\documentclass[preprint,authoryear]{elsarticle}
\usepackage{fullpage}
\usepackage{subcaption}
\usepackage{pifont}
\usepackage{latexsym}
\usepackage{mathrsfs}
\usepackage{amssymb,amsbsy}
\usepackage{amsfonts}
\usepackage{graphicx}
\usepackage[colorinlistoftodos]{todonotes}
\usepackage{varwidth}
\usepackage{lmodern}

\usepackage{mathtools}      
\usepackage[titletoc]{appendix}
\usepackage{ntheorem}
\usepackage{amssymb,amsbsy}
\usepackage{amsfonts}
\usepackage{graphicx}
\usepackage{tcolorbox}
\usepackage[colorinlistoftodos]{todonotes}
\usepackage{varwidth}
\usepackage{array}
\usepackage{makecell}
\usepackage[titletoc]{appendix}
\usepackage{titletoc}

\usepackage[fleqn]{nccmath}
\usepackage{algorithm}
\usepackage{algpseudocode}
\usepackage{longtable}

\usepackage{float}

\journal{}

\begin{document}

\begin{frontmatter}

\title{ALVEC: Auto-scaling by Lotka Volterra Elastic Cloud: A QoS aware Non Linear Dynamical Allocation Model}

\author[1]{Bidisha~Goswami}
\author[2]{Jyotirmoy~Sarkar}

\author[1]{Snehanshu~Saha}
\ead{snehanshusaha@pes.edu}
\author[3]{Saibal~Kar}
\author[1]{Poulami~Sarkar}

\cortext[cor1]{Corresponding author}

\address[1]{Department of Computer Science and Engineering, PESIT-BSC, Bangalore}
\address[2]{GE Healthcare, India.}
\address[3]{Center for Studies in Social Sciences, Calcutta, India and IZA Bonn.}

\begin{abstract}
Elasticity in resource allocation is still a relevant problem in cloud computing. There are many academic and white papers which have investigated the problem and offered solutions. Unfortunately, there is scant evidence of determining scaling quotient dynamically. Scaling is essential to maintaining elasticity in resource allocation. Elasticity is defined as the ability to adapt with the changing workloads by provisioning and de-provisioning Cloud resources. We propose ALVEC, a novel model of resource allocation in Cloud data centers, inspired by population dynamics and Mathematical Biology,  which addresses dynamic allocation by auto-tuning model parameters. The proposed model, governed by a coupled differential equation known as Lotka Volterra (LV), fares better in Service level agreement (SLA) management and Quality of Services (QoS). We show evidence of true elasticity, in theory and empirical comparisons. Additionally, ALVEC is able to predict the future load and allocate VM's accordingly. 
The proposed model, ALVEC is the first example of unsupervised resource allocation scheme.
\end{abstract}

\begin{keyword}
Cloud data centers, resource allocation, elasticity, Lotka Volterra (LV), population dynamics, Cloud Systems modelling, simulation, Resource allocation.
\end{keyword}


\end{frontmatter}

\section{Introduction}
Let us consider a hypothetical scenario in the Amazon rain forest where goats roam free without fear of being attacked or ambushed. Except natural death, the population doesn't diminish, in fact, is balanced by reproduction. The grassland may lose all the green since the goat population is not controlled. Whenever that happens, it is disastrous for goats as well since they'll have nothing left to eat. This may lead to migration and other critical consequences. On the contrary, if all goats are either killed or dead because of some natural calamity, the grassland is not consumed and for the lack of predators (goats) grass may grow in an uncontrolled fashion. Evidently, the balance between the two populations need to be maintained for a healthy ecosystem. Let us extrapolate this to a classical supply-demand scenario where regular and substantial supply of resources need to be fed to a stream of demands for jobs (not necessarily constant, may fluctuate with time). It is not impractical to associate the demand and supply with predator (goat) and prey (grass) population respectively. In turn, the predator-prey dynamics may be thought of as relevant interaction between the resources i.e. virtual machines (VM) as prey and demand i.e requested jobs (represented by cloudlets in cloudsim\footnote{\textbf{The cloudsim is a well known framework for modeling and simulating the cloud computing infrastructure and services. It is written in Java}}) as predators. It is well known that these terms, VM and jobs are integral to cloud computing \cite{Subhashini}, \cite{Beloglazov1}.
\par Cloud computing is an Internet-based computing. The computing paradigm revolves around providing shared resources and data to computers and other devices \cite{Xiangzhen}. However, instead of an apriori or ad-hoc distribution, the policy is implemented on demand. The shared pool of computing resources or prey as we may call those (VM), may be rapidly furnished and released to jobs or demands, conveniently re-coined as predators. This needs to be achieved with minimal effort and oversight. The cloud user may subscribe to these resources on short notice. The flexibility in pricing policy allows a subscriber to exploit these resources on pay-per-use/short term as well. Cloud computing allows the entrepreneur shun upfront infrastructure costs. Rather than mulling over infrastructure, he/she may focus on other operational and strategic initiatives. The fact that the unit cost of operating a server in large farms is relatively lower than small data-centers is an added incentive. Cloud provides virtual machines (prey) which accept the user requests (predators) and allocate the available physical resources accordingly. Cloud service provider acts as a broker between user requests and the cloud. The major challenges that confront these Cloud service providers are provisioning the Cloud resources in a dynamic environment without compromising the quality of service. Highly volatile nature of the demand of Cloud resources makes the chances of over-provisioning / under-provisioning of resources a common occurrence. Furthermore, maintaining competitive cost and pricing model adds to the complexity of the problem.
\par  Predator-Prey interaction forms the founding principle of population dynamics where the population are inter-dependent on each other. The study of population dynamics focuses on all the members of a single species who live together in the same habitat and are likely to interbreed, their unique physical distribution in time and space, growing or shrinking rate of population, etc. The predator-prey behavior signifies, if food is available in large quantity, then high food consumption increases the population of predator and large amount of prey consumption reduces the number of prey. At this point, because of scarcity of food available, number of predators may decrease. This is how the predator-prey model maintains both the populations dynamically. A similar kind of inter-dependency is observed between resources and the jobs. Let us assume the resources (VM's in CloudSim) as prey population and jobs as predators. When huge resources are available, Jobs can make use of sufficient options to avail the resources i.e VM's. However, when large amount of resources is consumed by jobs, the unavailability of resources (VM's) can deprive new jobs. This is the point where we face the challenge of elasticity triggering violation of SLA (Service Level Agreement), or additional resource provisioning which increases the cost. Moreover, there is always a threat of being penalized for frequent SLA violations in terms of either billing adjustments or even worse, migration of consumers to another provider.

A certain discourse in Mathematical Biology  models predator-prey behavior based on differential equations (see chapter 5 of \cite{Saha}). The first of those models is known as Lotka-Volterra (LV) population dynamics model \cite{LV1990}. The model discusses various situations that can take place based on the behavior of job and the available resources. The model has two fundamental equations which control the jobs and Virtual Machines(VM). Different parameters in the LV equations may control and interpret stability of the model. The LV model may fit a resource scheduling scenario, where different levels of dynamics between predator and prey population can be controlled by varying the parameters of the equations. According to LV, there must be provisions to control the system breakdown. The model boasts of a mathematical property, known as the limit cycle, which is described in contour portraits. Also known as phase portraits of the system, the graphical analysis help visualizing the balance between the two populations. Limit cycle describes a qualitative limit for the stability of a system whose parameters are differed such that the system grows out of stability. The difference acquired by these parameters is measured to describe the domain of stability. This may have direct application in understanding the stability of a web-server with incoming requests. Limit cycle of a system along with rate of incoming requests, can help us understand performance bounds of a system.\\
The dynamic interaction, observed in the natural habitat of  predator and the prey presents a compelling problem. Any service computing platform, distributed or otherwise needs to adapt to dynamically changing circumstances including adjusting to market volatility, fluctuating demand and supply etc. Some degree of autonomy must be granted to enable the components of service computing/IT enabled services to respond to system uncertainties. Therefore, the natural problems in the ecosystem and the challenge of balancing various issues may find a suitable sibling in cloud computing. The concepts imbibed from population biology need to be applied skillfully to address similar problems in cloud. To be more specific, solving the allocation problem in cloud data center by using the Lotka Volterra (LV) model is pertinent.
\section{Related Work}
In order to attain horizontal scaling, the user should define a fixed amount, say VM's to be allocated or deallocated. However, for vertical scaling, the same number  signifies the amount of resources(CPU, RAM) required to be added \cite{Lorido}. There are a couple of papers where upper and lower utilization threshold values of reactive scaling is the objective. Beloglazov et.al., introduces efficient adaptive threshold to meet the high level of SLA\cite{Beloglazov2}. Automated cloud-based scalability is a hot research topic in cloud computing. Fuzzy logic has been implemented in elasticity controller which enables  qualitative specification of elasticity rules \cite{Jamshidi}. Fuzzy logic in elasticity controller, utilized by Xu. et.al.,  has been used to learn the relationship between workload, resources and applied during resources allocation subsequently \cite{Xu}. Another approach in cloud controller is known as the black-box surrogate model, which evolves over time and uses machine learning to predict the performance \cite{Gambi}. 
Lim et. al.\cite{Lim} employed a linear equation to calculate the VM population in case of threshold violation (elasticity). The equation is heavily dependent on two parameters, actuator values and sensor measurement. CPU utilization is considered as sensor variable and actuator represents the number of virtual storage instances allocated as storage nodes. The relationship between workload and CPU utilization has been  established empirically. Whereas in Lotka-Volterra model, the major contributors in the equations are the number of virtual machines and the jobs. It is noteworthy that the biological model Lotka-Volterra \cite{Kolmogoroff} \cite{Keller} \cite{Goel} is a non-linear equation, which is reasonable as linearity may fail to explain the problem scenario. Chieu et.al.\cite{Chieu} have written a dynamic scaling algorithm for automated provisioning of virtual machine resources based on threshold number of active sessions. A hybrid controller, an amalgam of proactive and reactive controllers has been suggested by \cite{Urgaonkar}. Another work subscribes to the same concept and demonstrates the different possible scenarios of proactive elastic controller deployment in cloud incorporation with reactive elastic controller \cite{Eldin}.
\cite{Tesauro} demonstrates the strength of reinforcement learning in a sequential decision process, in which reinforcement learning (trains off-line) on data collected in combination with a queuing model policy controls the system. \cite{Arabnejad} proposed reinforcement learning with fuzzy logic to decide when to up /down scale instead of a predefined  threshold and the scaling action is a number from a fixed set {-2, -1, 0, +1, +2}. Though most of the authors consider two threshold values, upper and lower but  \cite{Hasan} have proposed 4 threshold values. ThrbU, is slightly below the upper threshold and ThroL is slightly above the lower threshold. A model-predictive algorithm defined by Roy et.al., is responsible for auto-scaling of resources.  A second order autoregressive moving average method (ARMA) is used to predict the workload and the optimization of the system behavior is achieved by minimizing various costs such as SLO violations, cost of leasing resources and reconfiguration cost\cite{Nilabja}. Waheed et.al. proposed a prototype based on reactive scaling, which continuously keeps monitoring the average response time. If the required response time is violated, it adds a VM \cite{Iqbal}, which is static way of deciding the scale number. SCADS has leveraged the utility function to scale-up and scale-down the storage resources dynamically and machine learning is utilized to predict the resource requirement of new queries before execution \cite{Armbrust}. \cite{Chaisiri} proposed an algorithm for optimal cloud resource provisioning using stochastic programming model to overcome the problem of On-demand cloud resource allocation plan. The author applied a decomposition algorithm to divide the actual optimization problem into multiple smaller problems such that these can be solved independently and in parallel. However the methodology has several complexities. The papers by \cite{Luck} and \cite{Kang} used agent technology to control dynamic environment like cloud. Singh et.al. have proposed a QoS  based resource provisioning and scheduling framework, where workloads are clustered using workload pattern and reclustered by k-means clustering algorithm to identify the Qos requirements. Different scheduling policies are employed to accomplish the scheduling task \cite{Singh}. Load balancing Ant Colony Optimization problem (LBACO) has been explored as a task scheduling policy which is a NP hard optimization problem. It incorporates the dynamic behavior of the cloud and balances the entire system \cite{ACO}.  Particle swarm optimization is another approach exploited in a previous paper, where computation cost and data transmission cost have been considered\cite{swarm}. Identical algorithm has been implemented in grid environment to achieve the optimized scheduling task \cite{Grid}. \cite{Varalakshmi} presented an optimal work-flow based scheduling (OWS) framework to identify a solution that can satisfy various user-desired QoS constraints, such as execution time. A comprehensive cost model, driven by partial utility, provided by client been been proposed. The cost model is effective in a scenario, where client is ready to accept a certain level of degradation \cite{Simao}.
\par Plenty of literature is available on task scheduling algorithm. \cite{Achar} presents a novel scheduling algorithm which utilizes the tree based data structure called Virtual Machine Tree (VMT) for efficient execution of the tasks. \citet{Vijayalakshmi} proposed a priority based task scheduling algorithm. In this algorithm, user tasks are prioritized and the task with highest priority will be assigned to a VM with highest processing power. \cite{genetictask} has enhanced the existing genetic algorithm in which the traditional algorithm Min-Min and Max-Min are merged with the standard Genetic algorithm. Min-Min and Max-Min are used to generate the initial population and it results in better solutions compared to standard Genetic  Algorithm in which initial population is generated randomly. In order to improve the energy efficiency and reduce carbon emission, a conceptual model and practical design guidelines for cloud resource management have been devised by \cite{buyya}. 
\par Lotka-Volterra model is widely used in the field of biological science especially to describe the population dynamics of two interacting species. Takeuchi et.al. considered the evolution system having predator prey deterministic systems denoted by Lotka-Volterra equations in random environment \cite{Takeuchi}. The periodic Lotka-Volterra predator-prey system  is investigated with impulsive effect \cite{Tang}. Chaos in three chain systems with LV model type interactions is showcased in another paper \cite{Chaos}. Nicola has made an attempt to establish a relationship between the LV model and predator-prey utility functions \cite{Nicola}.
\par The proposed Lotka-Volterra model, ALVEC has been integrated with standard task scheduling algorithm and improvement is observed by evaluating the performance on different QoS metrics. The intuition behind the LV time shared scheduling algorithm is to improve the performance and avoiding under-provisioning/ over-provisioning. Please note, load balance is not accommodated in ALVEC. However, LV timeshared algorithm is not dedicated to a particular environment, unlike some of the other work discussed in this section.

\section{Problem Statement:}
Achieving elasticity dynamically in allocating resources in cloud is a challenging problem. Though, there is some evidence of dynamic allocation of resources in cloud, but our proposed solution is first of its kind in this category with minimal oversight and control.
Lack of sophisticated models inspired us to propose a novel method of resource allocation in Cloud which addresses dynamic allocation and tunes the parameters of the model as per the on-demand service. However, design of such an automated strategy to scale resources up/down based on demand should not impact SLA management and Quality of services (QoS). The model should meet the standards of resource optimality, which signifies in improvements of quality metrics in cloud (simulated environment) such as vm utilization, SLA violation rate, average completion time etc. Additionally, the proposed model should be capable of predicting the future load and allocate the resources (VM) accordingly. 
\par \textbf{NOTE:} \textit{In the context of IoT/Edge computing, cloudlet it is a tiny datacenter. In the context of CloudSim, it is a class. Resource allocation is an abstract term, it can mean adding resources to a VM or add more VMs. We mean adding more VMs, rather than adding resources to a VM. Our theory and model are validated in a simulated environment, CloudSim.}
\section{Our Contribution:}
This paper introduces a new model inspired from population dynamics (LV) to control the system for maximizing the utilization of every resource. The goal is to handle resource under/over provisioning. 
\begin{itemize}
\item \textbf{Elasticity:}
Elasticity is defined as an ability to adapt with the changing workloads by provisioning and de-provisioning resources. The ability should be autonomic requiring minimal supervision. The proposed model implements elasticity by adjusting virtual machines in accordance with cloudlet demand. Most algorithms and strategies designed handle elasticity by increasing/decreasing VM's by one, by manual intervention or predefined rules. This is a supervised approach, even though not identified explicitly. We control the change in number of VM allocation/de-allocation exploiting the model dynamics proposed in our approach. Our approach is unsupervised and in contrast with the existing solution approaches. Unlike other well known approaches, agents or job managers are not required to allocate resources. Agents or Job managers are solely responsible to find out the number of resources, are required to allocate/deallocate  by understanding the demand of resources, therefore the entire process involves a considerable amount time and in some cases involve manual intervention. In the proposed case, the model decides the number of VMs considering the current resource demands and provides the input to the cloud VM allocation process. Auto-Scaling is equated to addition or removal of VM's in unsupervised fashion. Our model is adaptive, can auto correct allocation number based on demand in a completely unsupervised manner. Dynamic scaling is a well known feature of a commercial cloud provider but in many cases, the scaling number is being decided in a pre-defined manner. This is where our approach is distinct since we don't pre-determine the number of VMs while scaling. However, this didn't cause under-utilization of the VMs and as discussed in conclusion further (See \ref{vmelasticity}).
\item \textbf{Novelty of the model:} LV is extensively used in population biology. However, to the best of our knowledge, no application of the Lotka Volterra (LV) model in Cloud computing, in particular or communication networks, in general is found. This has the potential to set new baseline of research in Cloud Computing (See sections 5-8).

\item \textbf{Resource Optimality:} The proposed approach requires provisioning pooled resources. This may impact the performance metrics. However, we found that resource utilization is better compared to other algorithms in literature. At the same time, SLA violation minimization is ensured. The challenge in cloud resource optimization is scaling up or scaling down of resources based on dynamic need. An autonomous balancing model is proposed here which addresses equilibrium under volatility. 
\item \textbf{VM prediction based on population parameters:}
The proposed model specifies a upper and lower threshold. Threshold can be considered on any QoS metric (the model is not tightly coupled with any QoS metrics for the same) but  VM utilization and response time have been employed as threshold parameters. In the case of upper threshold violation by future response time prediction, new VM's need to be added to service to neutralize the situation. In the case of lower threshold violation, VM's need to be deallocated from the user service, as more than required number of VM's have been allocated to increase utilization of the resources (Details are discussed in section \ref{cloudsim}).
\item \textbf{Improvement in QoS metrics: make-span, response time and utilization:}
Make-span is the total duration between the job or service submission time and the completion time. Response time is computed by taking the sum of waiting time and execution time. Both are considered important quality metrics in analyzing performance of cloud data centers. The experiments based on our model show significant improvement in these metrics (detailed discussion may be found in section \ref{Technical}). 
\item \textbf{Parameter Tuning:} Parameter tuning implies controlling/ influencing the outcome of the parameters, which are nothing but VM's and jobs specified in LV model by changing the different coefficient parameters and satisfying the relevant conditions. The intention is to influence the values of the parameters as needed by manipulating the coefficients\footnote{There are total 4 coefficients in LV model $\alpha,\gamma,\beta,\delta$. For the sake of simplicity, $\beta,\delta$ are considered as constant and $\alpha,\gamma$ are allowed to fluctuate (increase or decrease) based on demand-service dynamics. Manipulating these aforementioned parameters acording to requirements such as prey increasing-predator decreasing, predator-prey stability is called parameter tuning.} of the parameters in the model. In this paper, we have exhibited how to cater to three different situations such as Prey Increasing-Predator Decreasing, Prey Decreasing-Predator Increasing and stability of Prey-Predator by tuning the parameters (See subsection \ref{parametertuning}). 
\item \textbf{Scheduling algorithm:}
The proposed scheduling algorithm mimicking the existing ecological model (non-linear in nature) address the dynamic nature adequately. The related papers show that the increase in the VM population is static and linear. The LV model decides the number of future VM allocation as per predicted need. Outside the purview of SLA, the model accommodates unanticipated load to be handled. 
\item \textbf{Application significance:} The proposed model is the first of its kind to balance the dynamics and auto-correct over/under utilization of resources. The applications are relevant in general cloud dynamics and data centers in particular.
\end{itemize}
Remainder of the paper is organized as follows.
Section \ref{Key Definitions} presents key definitions used from population biology and the relevant mathematical model. It is important to familiarize the readership with these definitions so that the mapping between LV and the Cloud problem be established clearly in section \ref{our model}. The analytical model presented in section \ref{our model} has to be solved numerically and qualitatively. These solutions along with the interpretation have been documented in sections \ref{solution proposed} and \ref{cloudsim} respectively. Section \ref{cloudsim} presents the simulation and outcome in detail. A numerical approach has bee discussed in section \ref{prediction} and various standard task scheduling algorithm are documented in section \ref{TaskScheduling}. This is followed by a detailed analysis of the benefits of service based outcomes in sections \ref{How LV} and \ref{Technical}.  We conclude the paper by discussing the advantages and pitfalls of our approach and background work.

\section{Key Definitions}\label{Key Definitions}
\begin{itemize}
\item Stability:
As per dynamic stability definition, the trajectories do not change too much under small perturbations. In cloud environment, stability is a condition where no significant changes occur in the VM or jobs. Hence, if perceptible change is not observed in the VM and job population (something that affects the gradient of both curves abruptly), there would not be any volatility in the model. This is the condition of stability. Steady state persists between VM and job when the stability condition is achieved. In other words, elasticity handling is not required at steady state.
\item Predator(Cloudlet):
Predator are the consumers of Prey. In this paper, we refer to job(cloudlets in cloudsim) as predator which consume VM's. In CloudSim, cloudlet is a class.
\item Prey:
Prey are consumed by the next layer of food-chain. We use a single layer of prey, which is Virtual Machine. \textit{Resource allocation, in our case, implies adding/removing VM's in an elastic manner.}
\item Qualitative theory of differential equations: In mathematics, the qualitative theory of differential equations studies the behavior of solutions without computing those. This is a visual exercise since finding solutions analytically is excruciatingly difficult, if not impossible. The paper studies equations \ref{preyEq} and \ref{predEq} which portray the dynamics of prey and predator population under different circumstances. 
\item Nullcline: In mathematical analysis, nullclines are encountered in a system of ordinary differential equations. The nullclines divide the phase portraits into regions. In this paper, the nullcline is equivalent to making equations \ref{preyEq} and \ref{predEq} zero. The subsequent calculations yield the conditions for stability in terms of the parameters of the LV model. This indicates the range of values of the parameters needed to be chosen such that elastic handling of resources is accomplished.
\item Equilibrium and stable condition: Equilibrium point is a constant solution to a differential equation. Fig. 2 explains a phase plot with a stationary point. The tentative point has no impact on existing VM or Cloudlet population. Controlling parameters helps ensure the system stability. 

\item Phase plane: Phase plane analysis is one of the most important techniques for studying the behavior of nonlinear systems. There is a direct method to show the existence of limit cycles. Fig. 2 is a phase-plane between Virtual Machines and jobs which explains the area where neither population creates impact on the other and shows the area where both the population are independent of each other. 

\end{itemize}
\begin{figure}[!ht]
  \centering
  \begin{minipage}[b]{0.45\textwidth}
   \includegraphics[width=\textwidth]{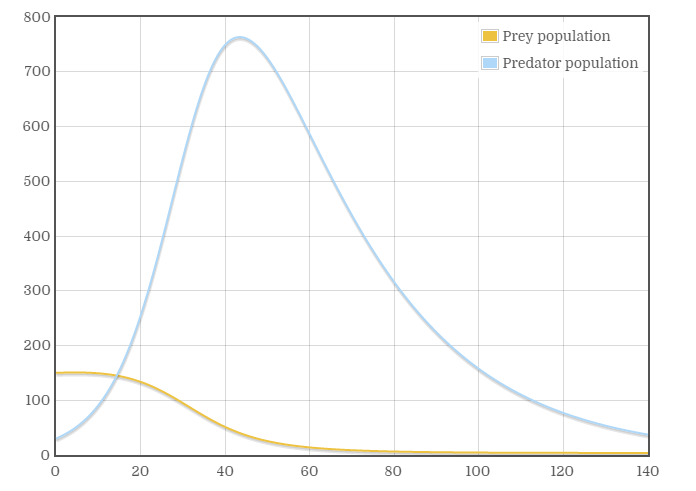}
\caption{Classical Lotka-Volterra plot for parameters computed: The graph consists of two different species of Food chain : Prey and Predator. In the proposed cloud model, Prey is Virtual Machine (VM) and Predator is jobs. The intersection point of predator and prey population is the NullCline point. If the VM population starts declining, then with a phase difference the jobs also decline for lack of resources.}
\centering\label{lkplot}
  \end{minipage}
  \hfill
  \begin{minipage}[b]{0.45\textwidth}
    \includegraphics[width=\textwidth]{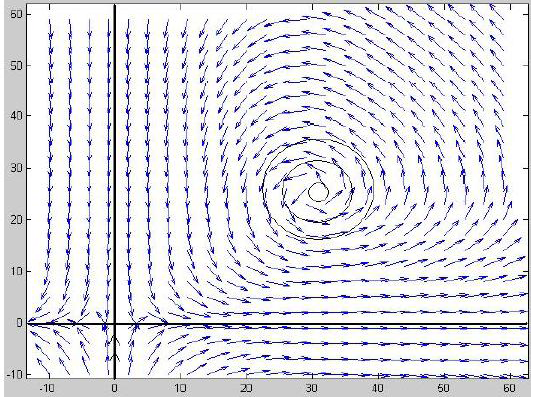}
\caption{This explains the stationary point $( \frac{\gamma}{\delta}, \frac{\alpha}{\beta})$. The stationary point has no impact on resource or job population. Controlling these parameters enables us to control the population dynamics and also control both Q and P populations. External management of these parameters determine the stability or instability of the system. Computation of stationery points may be found in Appendix A in additional file \cite{LV}.}
\centering\label{phaseplot}
  \end{minipage}
\end{figure}

\section{Our Model: Theory, Relevance and Applications}\label{our model}
\subsection{Relevance of Predator-Prey Model and Population Dynamics in Cloud Data Center}

The biological population dynamics model, Predator-Prey, is implemented to solve the problem. This model is germane to the different scenarios involving VM and job population. VM and jobs are be considered as the Prey and Predator respectively. The VM's can be consumed as long as enough number of VM's exist in the system or may be replenished (scaled up) to fulfill the need. Annihilation of prey population ensures similar consequences for the predator (please refer to the Amazon rain forest analogy in the introduction). In the same way, if all VM's are consumed or no VM is present in the system, the jobs (Predator here) gradually start loosing relevance which is equivalent to death or whole scale migration to another prey population. The herding tendency, though natural, increases the operational cost. The Predator-Prey model can be mathematically represented as VM-job model in the following manner:
\begin{ceqn}
\begin{align}\label{preyEq}
\frac{dP}{dt} = \alpha P - \beta PQ
\end{align}
\end{ceqn}
\begin{ceqn}
\begin{equation}\label{predEq}
\frac{dQ}{dt} = \delta PQ - \gamma Q
\end{equation}
\end{ceqn}
where, P is the number of VM's (Prey); Q is the number of jobs (Predators); \\
$ \frac{dP}{dt} $represents the growth rates of VM and $ \frac{dQ}{dt} $ represents growth rate of jobs over time;\\
$ \alpha $ is the upscaling  rate of VMs in the case of demand (jobs); \\
$ \beta $ is the allocation rate of VM  due to the incoming jobs;\\
$ \gamma $ is the completion rate of jobs; $ \delta $ is the job incoming rate into the system. To analyze the model in detail, the trend of $ P $ and the trend of $ Q $ need to be investigated. To make the dynamics stable, both the populations have to satisfy: $\frac{dP}{dt}=0 $ and $\frac{dQ}{dt}=0$ \\
\begin{ceqn}
\begin{equation}\label{stabilityEq}
\alpha - \beta Q = 0 , \delta P - \gamma = 0
\end{equation}
\end{ceqn}
Equation \ref{stabilityEq} evaluates a stationary point $( \frac{\gamma}{\delta}, \frac{\alpha}{\beta})$. Fig. \ref{phaseplot} represents the phase-portrait for the above dynamics. Notice from Fig. \ref{phaseplot} that all variations of population encircle around a stationary point.
\footnote{\textbf{In a cloud datacenter, it is not possible to control the inflow of jobs/ resources requests.But a cloud datacenter can maintain a pool of VMs to satisfy the incoming job requests.If there is a scarcity of prey(VM), the predator will migrate to another location(datacenter). The model never tries to manipulate the job number rather it suggests the possible VM number require to satisfy the current job requests}}

\subsection{Predator Prey Equilibrium and Regions}
Equations \ref{preyEq} and \ref{predEq} determine both the populations. Clearly, a greater number of VM's in the system is good for the model as multiple resources are present for consumption which makes it robust. Whereas, more population of jobs is a challenge as it may consume and thereby reduce the number of available VM's. To understand the stability, nullcline scenario of the predator-prey model should be discussed. In this model, two nullcline situations are possible, P-nullcline and Q-nullcline. P-nullcline is the set of points where $\frac{\partial P}{\partial t}=0$. Similarly, Q-nullcline is the set of points where $\frac{\partial Q}{\partial t}=0$. Now, by definition, equilibrium points are the locations, where growth rate of predator and prey both become zero. Hence, it can be said that P-nullcline is the location where growth rate of prey becomes zero. This signifies that prey population is neither increasing nor decreasing. On the other hand, the growth rate of predator becomes zero in the region of Q-nullcline. Apart from P-nullcline and Q-nullcline regions, the growth rate of Predator and Prey would be either positive or negative. Therefore the equilibrium points are located in the intersection of P-nullcline and Q-nullcline. Therefore, the neighborhood of P and Q nullclines are regions where the VM's and jobs do not fluctuate and elasticity management module (our hallmark contribution) is not invoked. However, in the other regions apart from the nullclines, elasticity management is a must!\\
The P-nullcline and Q-nullcline can be defined as below 
\begin{ceqn}
\begin{equation}\label{pnullclineEq}
\alpha P-\beta PQ=0,\\\hfill
\delta PQ- \gamma Q=0
\end{equation}
\end{ceqn}
The above equations can be rewritten as
\begin{ceqn}
\begin{equation}\label{qnullclineEq}
P(\alpha-\beta Q)=0, \vspace{2em}\\\hfill
Q(\delta P- \gamma)=0
\end{equation}
\end{ceqn}
From the above equations, it can be derived
$P=0$ or $\alpha-\beta Q=0$ 
and
$Q=0$ or $\delta P- \gamma=0$\\
The equilibrium points are $(0,0),(\frac{\gamma}{\delta},\frac{\alpha}{\beta})$. For simplifying the equation, we consider $\delta=1,\beta=1$, hence the co-ordinate points are $(0,0),(\gamma,\alpha)$ in the phase-plane.
The situation where stability can be achieved is described by
\begin{ceqn}
\begin{align}\label{newStab}
\alpha P=\gamma Q
\end{align}
\end{ceqn}
This can be derived from equations above.
As in stable situation, there is no growth for predator and prey.
Hence,
\begin{ceqn}
\begin{align*}
\alpha P-\beta PQ=0\\
=>\alpha P=\beta PQ
\end{align*}
\end{ceqn}
\begin{ceqn}
\begin{align*}
\delta PQ-\gamma Q=0\\
=>\gamma Q=\delta PQ
\end{align*}
\end{ceqn}
After considering $\beta=1,\delta=1$, the above equation can be rewritten as below
\begin{ceqn}
\begin{align*}
&=>\alpha P=\gamma Q
\end{align*}
\end{ceqn}
It is already proven that in stable situation, the value of $\alpha=Q,\gamma=P$
\begin{ceqn}
\begin{align*}
PQ=PQ
\end{align*}
\end{ceqn}
\begin{figure}[h!]
\centering
\includegraphics[width=0.6\textwidth]{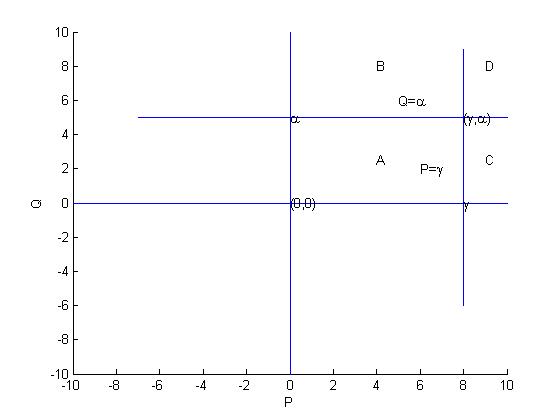}
\caption{represents the P-nullcline and Q-nullcline equations, \textbf{ where the x-axis depicts the prey P (jobs) and y-axis showcases} the predator Q (VM). The equilibrium points and different regions are visible in the figure. Two equations $\alpha=Q$ and $\gamma=P$ are plotted in the graph. The intersection of these two equations yields equilibrium points $(\gamma,\alpha)$. A is the region enclosed by $P=\gamma,Q=\alpha$, P, Q axis, where growth of P(Prey) is positive and the growth of Q(Predator) is negative. B is the region, which is the upper side of $Q=\alpha$ and left side of $P=\gamma$. In this region, the growth of P(Prey) is negative and the growth of Q is negative.
Region C is the right side of $P=\gamma$ and the lower side of $Q=\alpha$. The \textbf{growth rate of P is positive and the growth rate of Q is positive.
Region D defines the} upper side of $Q=\alpha$ and the right side of $P=\gamma$. The growth rate of P and Q are negative and positive respectively.}
\label{plotxyregion}
\centering
\end{figure}






\section{Solution to the proposed model}\label{solution proposed}
\begin{itemize}
\item The solution to the proposed system is critical in order to exploit the solution in the simulation and to explore QoS metrics. In this case, a closed form solution is the most convenient way of bringing out direct relations between the variables, predator (jobs) and prey (VM). Such direct relationship is often solicited since it explains the dynamics between the two key entities in Cloud computing. Regrettably, LV equations are inherently complex and do not admit of closed form, analytical solutions. Therefore, alternative methods to interpret and utilize the relationship between predator (jobs) and prey (VM) must be sought.
\item There are two ways to handle this. We use the qualitative theory to interpret the solutions and represent those in the phase plane (refer to definitions and relevant theory sections, section V). The representation of the solution qualitatively re-establishes our claim that the model is relevant in the context of \textbf{resource allocation and related issues in Cloud. However, in the absence of} explicit solutions, it is difficult to proceed further in the direction of exploiting the solutions in simulation and compute/tune parameters for performance enhancement (refer to QoS figs in discussion section).
\item We mitigate the problem by computing the solution to LV numerically. The numerical solution is central to our efforts in computing the parameters/coefficients in LV which further aids in accomplishing efficient VM allocation. This is accomplished by Runge Kutta and described in the next section.
\end{itemize}
\section{Numerical Solution of Lotka-Volterra}\label{prediction}
To retrieve result more accurately from LV model, we have employed the use of Runge kutta methods of fourth and fifth order, implemented by Fehlberge and denoted as RKF45. Numerical analysis to solve LV model is appreciated due to the inherent difficulty of solving the LV model analytically. We proceed in the following manner:\\
RKF45 produces an approximate solution in vector form $y_n$ by dividing the solution domain (Euclidean or Hilbert space, typically) into a set of discrete points. We begin with the initial data at time $t_0 = 0$ and estimate the
approximation solution at time $t_i=i*h$, i = 1, 2, ...n. The step size h is chosen suitably such that it is not too big or too small. We use RK4 and RK5 (Runge Kutta 4th order and 5th order respectively) at each step i to generate two different solutions and compare the proximity of the solutions thus generated. The approximation is acceptable within a certain tolerance as long as the difference between the two approximations doesn't exceed the predefined tolerance. The step size  may be modified to accommodate the tolerance criterion. However, we need to increase the step size if the two approximate solutions agree to more significant
digits than required. \textbf{Numerical methods are sensitive to approximation }and thus the following points must be stressed:
\begin{enumerate}
\item We use Taylor series expansion of the function around the iteration point at each step to approximate a function. This produces truncation error, large or small depending on the number of terms used in the expansion. If $h_n$ denotes the difference between $n+1^{th}$ and $n^{th}$ iteration, then a fourth order method produces an error of the form $Ch^5$ for some constant C. This means that a step size of magnitude $\frac{h_n}{2}$ shall reduce the error by a factor of $2^5 = 32$.
\item A 5th order Runge-Kutta method requires executing four function evaluations to obtain local truncation error of order 5. We observe, the numerical solution to the ordinary differential equation can be $5^{th}$ order accurate locally but may still not address the issue of global convergence adequately.
\item Roundoff error is inevitable. The estimate of VM's turns out to be a ballpark figure, precisely for this reason.
\item The population dynamics may deviate slightly from the standard assumption about the model for the above reasons.
\end{enumerate}
We adopted Runge-Kutta-Fehlberg 45 (RKF45) method, (ode 45 in Matlab) symbolized by function evaluations with an additional evaluation to accomplish 5th order accuracy. This  generates a local error of the order $h^6$, significantly small if $h$ is chosen to be small enough. Please note, $h$ is chosen to be between $0$ and $1$.
\par The parameters for simulation are computed using this method. ode45 of matlab (Appendix B in Additional File), which employs Runge-kutta method is used rigorously to derive the datasets  table \ref{table:2}, \ref{table:3} and \ref{table:4} corresponding to the three cases: Prey Increasing-Predator Decreasing, Prey-Predator stability and Prey Decreasing-Predator Increasing. 
\section{Task Scheduling Algorithms}\label{TaskScheduling}
We compared our model performance with Standard task scheduling algorithms in CloudSim later in the manuscript. We list those frequently used algorithms below:
\subsection{First Come First Serve }
This is one of the most simple algorithms and very easy to implement. The job/ task arriving first in the queue is assigned accordingly to a VM for execution. As it doesn't consider the execution time of the arrived task before allocation, sometimes it doesn't result in an efficient load balancing. A short job has to wait for longer time until a longer job finishes its execution. Therefore, it does not guarantee good response time. 
\subsection{Round-Robin Algorithm} It is a well known algorithm widely used in scheduling and load balancing. It selects the first VM randomly, assigns the tasks and selects the next VM/node in a circular manner (\cite{Round-Robin}).The advantage of the Round-Robin (RR) algorithm is it's simplicity. Sometimes, RR algorithm doesn't allocate the tasks to VM efficiently because it doesn't consider load, space, response time or any other parameter while allocating. Another variant of RR algorithm is weighted RR algorithm where each VM/node is assinged with a weight. The VM with more weight receives more tasks. If two VMs have equal weight, they both will be allocated with equal number of tasks. 
\subsection{Shortest Job First}
This scheduling algorithm selects the task having lowest execution time and assigns to a VM first. The job which has the highest execution time will be given the lowest priority. If two jobs demand equal execution time, it follows FCFS scheduling.
\subsection{Longest Job First} The job with the longest execution time is assigned to a VM. This is in stark contrast to SJF algorithm. SJF has disadvantages such as starvation, where a job with longest execution time waits for long time. If there is a flow of jobs which are shorter in execution time, then the longest job will not be assigned to any VM. To overcome this, LJF can be used in Cloud environment.
\subsection{Opportunistic Load Balancing Algorithm} Opportunistic Load Balancing (OLB) algorithm tries to keep the nodes busy irrespective of their current workloads. It assigns the task to a node in a random fashion. As it doesn't consider the current workload before assigning the task, sometimes it doesn't produce desired performance. 
\subsection{Min-min Load Balancing Algorithm} This is a static load balancing algorithm as it needs to know all relevant parameters before assigning the task to a node. It calculates the probable execution time and the completion time of all the tasks waiting in the queue. Then the task with minimum execution time is allocated to a node/VM, which requires minimum completion time. Therefore, tasks with maximum execution time has to wait until other tasks/jobs are assigned to the VMs. The completion time has to be updated when a task is assigned to a VM so that the task is removed from the meta task list. The entire process continues till the meta task list becomes empty. The Min-min algorithm  performs better than many other load balancing algorithms. But the algorithm needs to have knowledge of the execution time, completion time in advance before taking decision regarding allocation of the tasks.

\section{Simulation in CloudSim}\label{cloudsim}
For the lack of access to physical data centers, CloudSim offers a simulation framework for modeling  Computing infrastructures and services. Here, the biological model, Lotka-Volterra, explaining dynamic simulation, is implemented on the simulation platform (Refer section \ref{cloudsim}). In CloudSim, cloudlet is a class and is used to models jobs/demands in data centers. Resource allocation is equated to addition or removal of VM's. The quality metrics, which is being compared within different simulations, is Performance/Request Completion time, which is nothing but the time difference between the first time cloudlet request is submitted to the broker and the cloudlet completion time. We have kept the VM number constant for each data point across simulations and vary the cloudlet number, as more available VM will lead to better performance, which in turn disrupt the fair comparison. All jobs have been submitted dynamically. In almost every simulation, the jobs are dynamically submitted within a time frame, which is 1000 ms. Two situations are highlighted in this section. We consider a situation where the cloudlets for each simulation have been submitted in three batches. The CloudSim code has been modified to meet that objective.\footnote{We equate resource with VM and job with cloudlet thorough out section 8 and wherever simulation using cloudsim is discussed.}
Initial batch of VM and cloudlets are identical for each simulation to have a fair comparison. We have made use of two data centers in cloudsim, data center 1 and data center 2. Each data center has 2 host machines, having quad core and dual core processing capability respectively. Each host has $16384$ MB of RAM, $1$ GB of storage and bandwidth of $10000$ Kbps. The RAM size of VM is specified at 124 MB. As we wish to auto-scale from 100 VM's to 150 VM's, the RAM size of host machines is kept approximately 133 times that of the RAM size of a VM. Each VM has identical configuration. MIPS of VM is kept at $100$ while the MIPS of host machine is 240 times bigger of any VM. Bandwidth allowed to a VM is $100$ while a host machine enjoys $100$ times larger bandwidth. Each VM consumes a single CPU.

\subsection{ Case 1: Prey Increasing-Predator Decreasing}
This scenario may arise, when there is no VM available to provision or number of available VM is nearly 0 and requests for VM is rising. Such situation can be managed by reducing the number of cloudlets (either rejecting the incoming cloudlets or putting the cloudlets in queue) and increasing the number of VM's. LV model in such cases may suggest the required VM to mitigate the demands of cloudlets and the incoming cloudlets can be pushed to a queue till the data center is replenished with feasible VM population.
Say, we have 30 VM's available and at that moment the number of cloudlets is 50. Now, we would like to shoot up the number of VM's by increasing/deceasing the constants of the proposed model while the condition \(\gamma Q>\alpha P\),  mentioned in the algorithm \ref{algorithm1}, has to be met. \footnote{As we don't impose any control  on the incoming job requests, it will be more convenient to retrieve the required VM population from the model as per the cloudlet requests and letting the VM allocation process allocate the VMs from the pool. The model does not interfere in the allocation methodology, it only suggests the scaling number. VM is the only entity, which the data center has control of.} 
\begin{figure}[h!]
\centering
\includegraphics[width=0.6\textwidth]{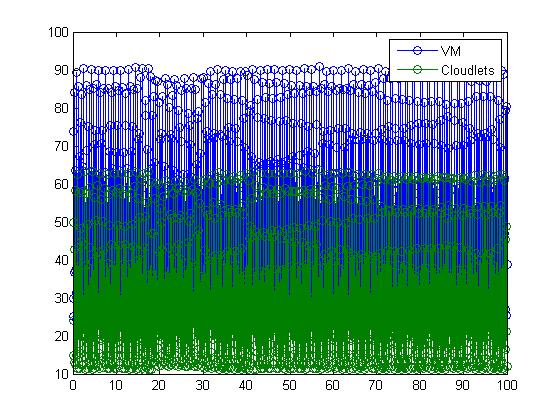}
\caption{Variation of cloudlets and VMs w.r.t time (Case 1). In the figure blue, green display the VM and cloudlets respectively. Y- axis represents the time-span, which is taken 0 to 100 duration. Each data point has been plotted at 0.01 time span interval. It is palpable from the figure that VM occupies the \textbf{upper part of the figure, whereas the cloudlets cover the lower part.} Hence, in maximum cases the values of VM are higher than cloudlets. The region, where VM and cloudlets values are overlapping lies within 60 to 35. The maximum value, which is belongs to VM lies in the region nearby 90, whereas the lowest value, belonging to cloudlets is nearby 10.}
\label{preyincre}
\centering

\end{figure}

The Lotka-Volterra model, depicted in the Fig. \ref{preyincre} following the equations:
\begin{ceqn}
\begin{align*}
\frac{\partial P}{\partial t}=30P-PQ; 
\frac{\partial Q}{\partial t}=-50P+PQ
\end{align*}
\end{ceqn}
where \(\alpha=30,\gamma=50,\beta=1,\delta=1\)\\

8 simulations have been performed where 2 simulation datasets are collected from Table \ref{table:2} and remaining datasets are randomly chosen.
VM number is kept constant across the simulations. Initial data points remain unchanged for simulations to have a fair comparison. Lowest avg completion time is 400, which is visible in simulation 1. Though the difference among various average completion time is few milliseconds, the performance of simulation 1, which is derived from model is better than others. Here, dynamic influx of cloudlets within 1000 ms is considered, i.e. within 1000 ms, all the cloudlet requests arrive at the data center. 
\begin{table}[!ht]
    
    \begin{minipage}{.45\linewidth}
    \centering
      \begin{tabular}{c c c } 
\hline 
Time&VM&cloudlets\\ 
\hline
0 & 30 & 50\\ 
0.1 & 73.817541 & 14.87\\ 
0.2 & 25.19 & 23.96\\ 
0.3 & 84 & 42.70\\ 
0.4 & 36.78 & 12.97\\ 
0.5 & 37.24 & 58.28\\ 
0.6 & 63.62 & 12.56\\ 
0.7 & 24.35 & 30.050\\ 
0.8 & 89.46 & 30.17\\ 
0.9 & 31.52 & 14.92\\ 
1.0 & 49.80 & 62.49\\ 
1.1 & 53.56 & 11.46\\ 
1.2 & 25.13 & 38.34\\ 
1.4 & 85.68 & 20.72\\ 
1.5 & 27.62 & 18.14\\ 
1.6 & 67.18 & 57.75\\ 
1.7 & 44.61 & 11.51\\
1.8 & 28.59& 48.25\\
1.9 & 76.06& 15.35\\
2.0 & 25.04& 22.87\\

\hline
\end{tabular}
\caption{This table demonstrates a scenario (Case 1), where it is required to increase the Prey(VM) number. In the table, time-span from 0 to 2.0 has been taken for better understanding of how the model works. Initially, the predator and the prey numbers are taken as 30 and 50 respectively. Time-span is displayed in the table for every 0.1 interval. As the intention was to increase the number of prey from 30, in the next immediate time-span it can be noticed that the prey number surges to 73, a two fold jump from the initial value. Apart from a few occurrences, through out the period till 2.0, the number of prey is higher than initial value. In the case of predators, the number of predators are less than the initial value 50. Only one occurrence at time-span 1.0, the predator population is more than the initial value. The prey-predator numbers in the table and the figure will be different, if any other initial values are considered. As the proposed model is not a linear function, there is no pattern visible in the prey-predator numbers. }\label{table:2}
    \end{minipage}%
    \hfill
    \begin{minipage}{.5\linewidth}
    \centering
      \begin{tabular}{c c c c } 
\hline 
Simulation No&VM&cloudlets&Avg Request Completion time\\ 
\hline
1 & 30 &50&499.76\\ 
1 & 36& 12 & 400\\ 
1 & 76& 15 & 400\\
2 & 30 &50&548.24\\ 
2 & 36& 112 & 536.67\\ 
2 & 76& 115 & 540.62\\ 
3 & 30 &50&531.92\\ 
3 & 36& 8 & 506.25\\ 
3 & 76& 115 & 543.64\\
4 & 30 &50&513\\ 
4 & 36& 200 & 530\\ 
4 & 76& 5 & 487\\ 
5 & 30 &50&493.28\\ 
5 & 85&20 & 400\\ 
5 & 44& 11 & 400\\ 
6 & 30 &50&514.1188\\ 
6 & 85&200 & 686.6909\\ 
6 & 44& 150 & 692.715\\
7 & 30 &50&543.08\\ 
7 & 85&10 & 400\\ 
7 & 44& 110 & 400\\
8 & 30 &50&512.52\\ 
8 & 85&110 & 479.44\\ 
8 & 44& 8 & 552.5\\
\hline
\end{tabular}
\caption{This table represents the simulations where Cloudlets (Predator) need to decrease and VM (Prey) number is supposed to increase (Case 1). Total 8 simulations have been performed. Out of these 8 simulations, for 2 simulations (Simulation No 1 and 5) data points are taken from LV model, whereas rest of the simulations consist of random data points generated in controlled manner. The average request completion time is calculated for each VM-Cloudlet pair. We have kept the VM number constant for each data point across simulations while varying the cloudlet number (since more available VMs will lead to better performance, which in turn disrupt the fair comparison). All cloudlets have been submitted dynamically}\label{simulation9}\footnote{The random data points are generated in a controlled manner. If we observe simulation 3, 2nd data point corresponding to a cloudlet is lower but 3rd data point is higher than the first simulation. In case of 2nd simulation, 2nd and 3rd  data point cloudlets numbers are greater than the first one.} 
    \end{minipage} 
\end{table}
\begin{table}[htb]

\begin{center}

\end{center}
\end{table}

\begin{table}
\begin{center}

\end{center}
\end{table}
The simulation 1 dataset is part of table \ref{table:2}, which is the master dataset derived from the proposed model.

Table \ref{simulation9} contains the average request completion time of Simulation 2. Here the initial data point, which is nothing but the first batch of Cloudlet submission is same as 1st simulation. But for the 2nd and 3rd data points, cloudlets are higher than the 1st simulation.
The simulation 3 presents another random dataset, where the 2nd data point is lower and 3rd data point is higher in comparison than Simulation 1.
The simulation 4 is the reverse of simulation 3 as the 3rd data point cloudlet number is higher than simulation 1 and the 2nd data point is higher than the 2nd data point of simulation 1.
Simulation 5 dataset belongs to the master dataset in table \ref{table:2}. Simulations 6, 7 and 8 are performed in random datasets which have been generated based on simulation 5 in a controlled manner.
The 2nd and 3rd data points of simulation 6 are higher than 5th simulation which is originated from proposed model.
Simulation 7 delivers almost the same performance as simulation 5. The 2nd data point is lower and 3rd is higher than the corresponding data points in simulation 5.
In the case of simulation 8, the 2nd data point is higher and the 3rd data point is lower than corresponding data point in simulation 1.\footnote{We equate resource with VM and job with cloudlet throught section 8 and wherever simulation using cloudsim is discuused.} We conclude from simulations 5, 6, 7 and 8 that the average completion time of simulation 5 is better than others.

\subsection{Case 2 :Prey-Predator stability}
This section highlights the situation where data center has achieved its maximum VM utilization target. Therefore, same number of VM and jobs need to be maintained afterwards (no volatility). The growth and decay rates of VM and cloudlets are 0. Stability implies no change in VM and cloudlet population as time passes. Same population of VM and cloudlet needs to be maintained once the desired utilization is reached. This situation is applicable if it is possible to maintain the same VM population for a certain period of time and the incoming cloudlets' requests do not fall below the expected number. Every data center might have a utilization threshold, beyond which it does not intend to stretch. It can be 80\% of total VM utilization or can be any number based on their business model and other criteria. The proof of stability in the proposed model is presented mathematically in Appendix A in Additional File \cite{LV}. To keep the same number of VM and cloudlets, the acceptable range of $\alpha$, $\gamma$ parameter values of Lotka-Volterra model has also been elaborated in Appendix in Additional file.
\begin{figure}[h!]
\centering
\includegraphics[width=0.6\textwidth]{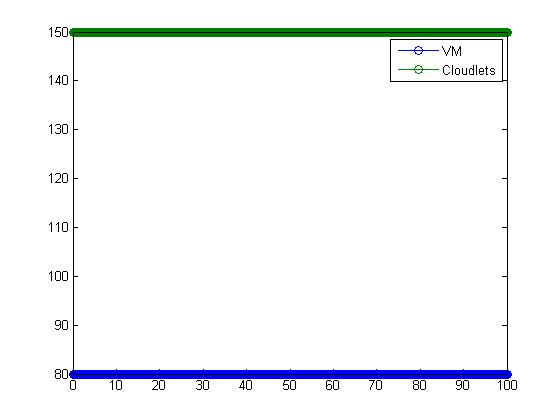}
\caption{ Illustration of variation of cloudlets and VMs wrt time in stable situation (Case 2). The X axis represents the VM, cloudlets number, whereas the Y axis represents time span. Blue color depicts VM and green color represents cloudlets. It is observed from the figure that there is no change in the number of predator or prey throughout the time period (0-100). VM and cloudlets maintain the same initial values, which are 80 and 150 respectively from time 0 to 100.}
\label{preystable}
\centering

\end{figure}

The Lotka-Volterra model, which is plotted in the Fig. \ref{preystable} is as follows:
\begin{ceqn}
\begin{align*}
\frac{\partial P}{\partial t}=150P-PQ\\
\frac{\partial Q}{\partial t}=-80P+PQ
\end{align*}
\end{ceqn}
Where \(\alpha=150,\gamma=80,\beta=1,\delta=1\)\\
The condition which needs to be satisfied to reach stable situation is: \(\gamma Q=\alpha P\), where $\gamma=P$  and $\alpha=P$.

\begin{table}[htb]

\begin{center}

\begin{tabular}{c c c } 
\hline 
Time&VM&cloudlets\\ 
\hline
0 & 80 & 150\\ 
0.1 & 80 & 150\\ 
0.2 & 80 & 150\\ 
0.3 & 80 & 150\\ 
0.4 & 80 & 150\\ 
0.5 & 80 & 150\\ 
0.6 & 80 & 150\\ 
0.7 & 80 & 150\\ 
0.8 & 80 & 150\\ 
0.9 & 80 & 150\\ 
1.0 & 80 & 150\\ 
1.1 & 80 & 150\\ 
1.2 & 80 & 150\\ 
1.4 & 80 & 150\\ 
1.5 & 80 & 150\\ 
1.6 & 80 & 150\\ 
1.7 & 80 & 150\\
1.8 & 80& 150\\
1.9 & 80& 150\\
2.0 & 80& 150\\

\hline
\end{tabular}
\caption{Predator Prey stability is the scenario (Case 2) where the same VM(prey) and cloudlets(Predator) numbers need to be maintained. The table \ref{table:3} displays the predator, prey numbers at each time point, which are collected after 0.1 time interval. The table also supports the conclusion drawn from the fig \ref{preystable} that there is no change in VM, cloudlets number as time passes from 0 to 100.}\label{table:3}
\end{center}
\end{table}

\subsection{Case 3: Prey decreases-Predator Increases }
This scenario may arise when VM number reaches maximum available capacity and there is a need to allocate more VMs to incoming cloudlets to improve utilization. Such a situation, where data center needs to concentrate on provision of idle VMs, requires to decrease available VM (prey) and increase cloudlets number(Predator). Available VM number decreases as it is being provisioned to different incoming requests. Number of Cloudlets rises because more VMs are available and ready to serve incoming requests. As the dynamics keep changing, the LV model plays a crucial role by suggesting the available VM pool to be dropped from the current VM pool.\\ 
\begin{figure}[h!]
\centering
\includegraphics[width=0.6\textwidth]{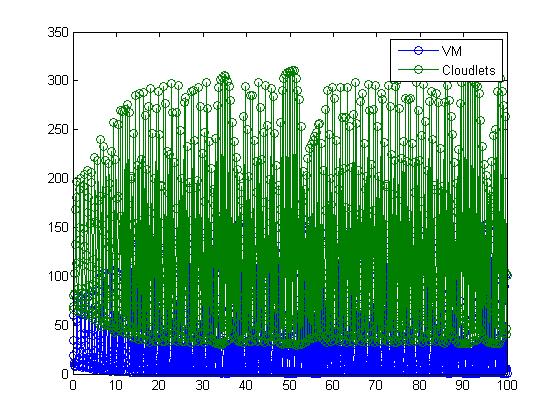}
\caption{Variation of cloudlets and VMs wrt time, when prey number is required to decrease (Case 3). X axis represents Time-span and Y axis represents VM and cloudlets number. The values are spread out between 0 to 350. The blue region, belonging to VM occupies the lower part of the figure, whereas the upper region is covered by green, which signifies cloudlets. There is a significant region overlapped by VM and cloudlets but maximum places of the figure are free from overlapping. The maximum value in the figure, which is attained by the cloudlets, is near by the region 300. Minimum value, which is belongs to VM is near by 0.}
\label{preydecrease}
\centering

\end{figure}
According to the algorithm, the condition \(\alpha P>\gamma Q\) has to be met, where \(\alpha>Q, \gamma<P\)
The Lotka-Volterra model, which is plotted in the Fig. \ref{preydecrease} is the following:
\begin{ceqn}
\begin{align*}
\frac{\partial P}{\partial t}=120P-PQ\\
\frac{\partial Q}{\partial t}=-30P+PQ
\end{align*} 
\end{ceqn}
Where \(\alpha=120,\gamma=30,\beta=1,\delta=1\)\\

\begin{table}[!ht]
    
    \begin{minipage}{.45\linewidth}
    \centering
      \begin{tabular}{c c c } 
\hline 
Time&VM&cloudlets\\ 
\hline
0 & 60 &80\\ 
0.1 & 34.73 & 66.19\\ 
0.2 & 18.97 & 69.16\\ 
0.3 & 11.40 & 82.47\\ 
0.4 & 8.33 & 103.14\\ 
0.5 & 7.99 & 132.47\\ 
0.6 & 11.21 & 168.11\\ 
0.7 & 23.44 & 197.33\\ 
0.8 & 55.89 & 180.68\\ 
0.9 & 76.83 & 112.69\\ 
1.0 & 53.72 & 72.76\\ 
1.1 & 28.69 & 64.30\\ 
1.2 & 15.14 & 71.04\\ 
1.4 & 9.29 & 87.74\\ 
1.5 & 7.23 & 114.07\\ 
1.6 & 8.34 & 150.29\\ 
1.7 & 15.28 & 189.09\\
1.8 & 40.13& 200.55\\
1.9 & 77.98& 137.38\\
2.0 & 64.06& 78.36\\

\hline
\end{tabular}
\caption{This table captures the situation where the VM(Prey) needs to reduce but cloudlets(Predator) number is required to increase (Case 3). The table  displays a few data points used to plot the figure. The initial VM, cloudlets values are 60,80. Except a few, all the VM values are less than initial VM value. In the case of cloudlets, there are a few occurrences, where cloulets values are less than initial value but maximum cloudlets values are higher than initial cloudlet value. }\label{table:4}
    \end{minipage}%
    \hfill
    \begin{minipage}{.5\linewidth}
      \centering
        \begin{tabular}{c c c c } 
\hline 
Simulation No&VM&cloudlets&Avg Request Completion time\\ 
\hline
1 & 60 &80&440.25\\ 
1 & 11 & 82 & 2832.10\\ 
1 & 8 & 103 & 3165.24\\ 
2 & 60 &80&435.95\\ 
2 & 11& 182 & 5751.57\\ 
2 & 8& 203 & 6505.15\\ 
3 & 60 &80&432.45\\ 
3 & 11& 182 &3591.95\\ 
3 & 8& 43 & 3636.95\\
4 & 60 &80&425.1\\ 
4 & 11& 52 & 2626.45\\ 
4 & 8& 143 & 3310.23\\
5 & 60 &80&445.90\\ 
5 & 55& 180 & 1028.18\\ 
5 & 9& 87 & 1074.31\\
6 & 60 &80&485.1785\\ 
6 & 55& 300 & 2672.91\\ 
6 & 9& 245 & 2655.57\\ 
7 & 60 &80&452\\ 
7 & 55& 90 & 1071.68\\ 
7 & 9& 175 & 1066.44\\ 
8 & 60 &80&462.026\\ 
8 & 55& 350 & 1570.62\\ 
8 & 9& 15 & 1804.732\\

\hline
\end{tabular}
\caption{This is the simulation scenario, where it is required to increase cloudlets (Predator) and to reduce VM (Prey) number (Case 3). Simulation No 1 and 5 are derived from the LV model and data points for remaining simulations are randomly generated but in a controlled manner }\label{simulation1}
    \end{minipage} 
\end{table}

A total of 8 simulations are conducted to calculate average completion time for each batch. Out of these 8 simulations, 2 simulations are executed using datasets, which belong to Table \ref{table:4}. For comparison purpose, 6 simulations are done over random datasets but in a controlled manner. The Simulation 1 dataset is derived from the proposed model and average Cloud request completions are calculated from Cloudsim. Simulations 2, 3 and 4 are performed on the random datasets to compare performance with simulation 1.



The difference of Simulation 2 with the first simulation is that the 2nd and 3rd data point cloudlet numbers are greater than the first one. The 2nd data point Cloudlet number of simulation 3 is higher but 3rd data point Cloudlet number is lower than the first simulation.


In simulation 4, 2nd data point Cloudlet number is lower whereas 3rd data point Cloudlet number is higher in comparison to the first simulation. The simulation 5 dataset is derived from predator-prey model (ALVEC). In simulation 6 dataset, the 2nd and 3rd data points are higher than the 5th simulation. In simulation 7, the 3rd data point performance is better than the 5th simulation, which is derived from our model. In simulation 7, the 2nd data point is lower whereas the 3rd data point is higher in comparison to the 5th simulation. In the case of simulation 8, 3rd data point cloudlet number is lower than 5th simulation and 2nd data point is higher than the corresponding point in 5th simulation. The 1st and 5th simulation (using dataset derived from our model) performed better than other random data sets. We noticed one exception in simulation 4,  where data point 2 needs lesser average completion time in comparison to simulation 1 (it has same VM number, 11 but lesser number of cloudlets). Simulations 6, 7 and 8 are random datasets in comparison to the derived dataset from simulation 5. Hence, we conclude that the dataset derived from ALVEC, performed better than random datasets (which follows no pattern).

The stable scenario is achieved, when \(\alpha=Q,\gamma=P\) condition satisfies. Other two scenarios discussed above, can be achieved by fluctuating the $\alpha,\beta$ values from the stable situation. The difference between the \(\alpha P\) and \(\gamma Q\) determines the behavior of the data points in the table and the figure. Therefore, this model exhibits an advantage, where the constants for the predator and prey can be chosen to decide the expected behavior.\\


\subsection{The modeling approach in CloudSim}
CloudSim is a framework for modeling and simulation of Cloud computing infrastructures and services. \footnote{We equate resource with VM and job with cloudlet throught section 8 and wherever simulation using cloudsim is discuused.} An acceptable  ecosystem of Cloud environment satisfying increasing demand for energy-efficient IT technologies, expects timely, repeatable, and reliable methodologies for evaluation of algorithms, applications, and policies before actual development of Cloud products. Utilization of real testbeds makes the reproduction of results an extremely difficult undertaking. Surrogate approaches need to be leveraged for testing and experimentation facilitating  the development of new Cloud technologies. However, simulation tools can be effectively exploited to evaluating the hypothesis for software development apriori. This has to be accomplished in a reproducibility- friendly environment. CloudSim is one such tool used for our simulation in two different ways and comparison of experimental results. \\

The initial approach is to allocate all the resources statically at the beginning of simulation. When the resources are allocated statically at the beginning of simulation, it results in over / under utilization and over / under provisioning of resources. Over-provisioning of resources occurs when the user requests gets surplus resources than demand. Under-provisioning of resources occurs when the user requests are assigned with fewer number of resources than the demand. Both over-provisioning and under-provisioning of resources result in poor optimization of resource allocation.  \\
\\
Next, we dynamically add the resources on-demand. Adding resources dynamically into the system avoids over / under provisioning of resources. Here the dynamic simulation model is compared with a biological model called Lotka-Volterra. 

The resources on CloudSim compared with Lotka-Volterra model are described as:  
\begin{itemize}
\item {P is the number of Virtual Machines (Prey)}
\item {Q is the number of cloudlets (Predators) where Cloudlet specifies the user request}
\item {$ \alpha $ is birth rate of Virtual machines in the absence of predation by cloudlets}
\item {$ \beta $ is death rate of Virtual machines due to predation}
\item {$ \gamma $ is natural death rate of cloudlets in the absence of Virtual Machines}
\item{$ \delta $ is reproducing rate of cloudlets }
\end{itemize}

The simulation model is used to compute the parameters of Lotka-Volterra model. These parameters are used to control the system.\\
\subsection{Resource Allocation algorithm using Predator Prey}\label{resourceallocation}
Cloud computing provisions resources on the basis of demand. One of the major aspects of Cloud computing is that it allows to scale up and scale down resource allocation based on needs. Predator-Prey model, Lotka- Volterra, can be employed to understand the behavior of need based resource allocation. Cloud computing has been built upon virtualization  and distributed computing to maximize resource utilization. Here, resources can be considered as prey and individual requests as predator. The objective is to establish that resources(VM) and requests(cloudlets) follow the Lotka-Volterra, Predator-Prey relationship.\\

\begin{algorithm}
\caption{Lotka-Volterra algorithm in Cloud Dynamics}\label{alg:lotka-volterra}
\label{algorithm1}
\begin{algorithmic}[1]
\Procedure{Lotka-Volterra}{p,q}
\Comment{p is prey(VM), q is predator(Cloudlet)}
\State $p\gets VMs$ \Comment{Initialize VM} 
\State $q\gets cloudlets$ \Comment{Initialize cloudlets}
\While{$ VM=0 $}\label{VMzero}
\While {$ ( \gamma \geq P  ) and (Q \geq \alpha) $} 
\State $ \gamma \gets \gamma +\epsilon $ \Comment{$\epsilon$ is infinite small number}
\State $\gamma Q \geq \alpha P$ 
\EndWhile\label{euclidendwhile}
\EndWhile\label{VMZero}
\While {$ VM \gets maxVM and cloudlets \neq 0$}
\While {$ ( \alpha > Q ) and (\gamma < P)$}
\State $ \alpha \gets \alpha +\epsilon $ \Comment{$\epsilon$ is infinite small number}
\State $\alpha P \geq \gamma Q$ 
\EndWhile\label{}
\EndWhile\label{maxVM}
\State \textbf{return} 
\EndProcedure
\end{algorithmic}
\end{algorithm}
\underline{\textbf{Algorithm Explanation:}}
The algorithm starts with the initialization of prey(VM) and predator(Cloudlet). In a Cloud data center, if such situation occurs when no VM is available for allocation to newly arrived jobs, then the value of \(\gamma\) needs to increase in such a way that it satisfies \(\gamma Q>\alpha P\) where \(\gamma >P,\alpha<Q\). Therefore the VM number increases and cloudlets number decreases. If the VM number is near the maximum available VM and cloudlets are available then the value of \(\alpha\) (weight of P) needs to increase so that \(\alpha P>\gamma Q\) satisfies, where \(\alpha>Q and \gamma<P\). Hence Q resources (cloudlets) in the system will increase and P (available VM) will decrease. In that case, VM attains the maximum utilization level and needs to maintain the same VM and cloudlets numbers. \(\gamma Q=\alpha P\) condition needs to be met, where \(\gamma=P, \alpha=Q\). Considering all the scenarios \(\beta,\delta=1\).

\section{How is LV helping in achieving what was not accomplished before? The Benefit Analysis}\label{How LV}
Ecological balance is one of the major areas of study for an ecologist. This model is important for the continued survival and existence of organisms without compromising the stability of the environment. As explained in \cite{Scott2010}, the systems are complex. The model describes a hierarchal structure of food chain and describes how every layer of predators have significant importance. Removal of any layer of predator challenges ecological stability by regulating the impacts of grazing. This ensures the overall productivity of the following layer of animals. Lotka-Volterra\cite{LV1990} is one of the most discussed model in food-chain system which describes the dynamics between any two corresponding layers of predator-prey relation. In service computing like Cloud computing, users can be considered as a predators. The user demands computing as a service and consumes the resources that Cloud provides. The Cloud resources are prey, which is consumed by the higher layer of food-chain, i.e users.
The model proposed in paper \cite{Goswami} based on the dynamic interaction between the predator and the prey. A multi-agent model was proposed to control the heterogeneous and volatile demand handling environment like Cloud. To address the volatility, some degree of autonomy is needed to enable the components to respond to dynamically changing circumstances. To address the above mentioned scenario, an elastic, autonomous and balancing model is needed to address the equilibrium under volatility. The proposed model addresses all the qualitative parameters with Lotka-Volterra model mimicking the ecological balance in ecology.\\

The paper uses LV( Lotka-Volterra) model to address more than one issue. These are parameter tuning, elasticity in VM, an improved Timeshared algorithm, improvement in QoS, reduction in SLA Violation and predictive Analysis for VM allocation. The following section contains details with explanation and experimental results. 
\section{Model Implementation and Outcome: Technical Discussion}\label{Technical}

\subsection{Parameter Tuning}\label{parametertuning}
Lotka-Volterra model can give a different direction regarding resources provisioning and de-provisioning dynamically as per workload changes. Lotka-Volterra will be very efficient to predict the number of virtual machines based on the number of incoming job requests and VM number when workload changes as per demand.

\begin{algorithm}
\caption{Scaling by Parameter Tuning}\label{alg:lv-parameter tuning}\label{paramtertuning}
\begin{algorithmic}[1]
\Procedure{LV-Parameter-Tuning}{}\Comment{}
\State $maxT\gets MaximumThreshold$
\State $minT\gets MinimumThreshold$
\State $VM\gets Number-of-VMs-in-present-VM-pool$
\Comment{Initialize VM} 
\State $T\gets Time$
\While{$ T \neq 0 $}

\While {$ CPU-Utilization > maxT$} \Comment{Trigger LV, VM increasing cloudlets decreasing}
\State $ (new-VM > allocated-VM ) and ( new-VM < allocated-VM + VM $ \Comment{}
\State $VM \gets VM + additional-VM-in-pool$ 
\EndWhile\label{maximum Threshold}

\While {$ CPU-Utilization < minT$} \Comment{Trigger LV, VM decreasing cloudlets increasing}
\State $ (new-VM < allocated-VM ) and ( new-VM > 0) $ \Comment{}
\State $VM \gets VM - LV-generated-number$ 
\EndWhile\label{minimum Threshold}

\State $ T \gets T-1 $
\EndWhile\label{Timevalidity}
\State \textbf{return} 
\EndProcedure
\end{algorithmic}
\end{algorithm}

\underline{\textbf{Algorithm Explanation}}
\begin{itemize}
\item Define maxThreshold and minThreshold of VMs and initialize VM pool.
\item Calculate VM utilization for every particular time interval.
\item If CPU utilization $ > $ maxThreshold.
\item Trigger Lotka-Volterra (Prey Increasing and Predator decreasing situation). Lotka-volterra  returns a set of VM number, cloudlets number.
\item  From this set select the particular VM number which is $ > $ current allocated VM and $ < $ current allocated VM $ + $ VMs in pool.
\item Add the additional VM from  VM pool.
\item if CPU utilization $ < $ minThreshold
\item Trigger LK (Prey Decreasing and Predator Increasing situation). select the VM number <currentOnlineVmnumber,  and VM number $ >0 $.
\item  Deallocate the VMs based on LK generated number and returned to the VM pool.

\end{itemize}
The parameter which is used as a criteria of provisioning and de-provisioning of VMs is utilization. Say a Cloud provider has decided to implement a monitoring algorithm, where the maxThreshold and minThreshold are defined as 80\% and 20\%. Every 30 seconds the algorithm will keep checking the VMs average utilization by using the formula given below \cite{Aslanpour}.\\

\begin{ceqn}
\begin{equation}\label{equti}
VM utilization=\frac{\sum_{I=1}^{currentcloudlets}CloudletLength(i)*CloudletPEs(i)}{PEs*MIPS}
\end{equation}
\end{ceqn}
\begin{ceqn}
\begin{equation}
\label{eqallvmuti}
VMs avg. Utilization=\frac{\sum_{i}^{OnlineVms}VM_{i}Utilization}{Online VMs}
\end{equation}
\end{ceqn}
In \ref{equti} currentcloudlets is nothing but the number of cloudlets arriving for a particular VM. PE is the number of processors in the VM. MIPS is the processing power of each processor core. CloudletLength is the number of instructions to be run. We have considered all the cloudlets as homogeneous (number of instruction or cloudlet length is fixed for every cloudlet). CloudletPE is the number of processors required by the Cloudlet request.
In equation \ref{eqallvmuti} OnlineVMs is the total number of VMs allocated for execution and $VM_{i}Utilization$ is the utilization of a single VM, calculated using \ref{equti}. VM's average utilization is compared with maxThreshold and minThreshold. If it violates the maxThreshold, it means VMs are over occupied and the number of VMs are not adequate to meet the spike of demand at that point of time. Hence, it is required to add more VMs from VM pool to serve all the incoming cloudlet requests without SLA violation and reducing the overhead on individual VM. But the question is how many VMs are needed to be pulled from the pool. Here the Lotka-Volterra algorithm plays its role by providing the VM number based on currently allocated VMs and total cloudlet number, executing in different VMs. In this particular scenario, Prey increasing and Predator decreasing condition is suitable as we need to increase online VM number. This monitoring algorithm never controls the incoming cloudlet number. If it does not satisfy the minThreshold criteria, it signifies that more than required VMs are allocated to the incoming cloudlets and VMs are under utilized. The number of VM needed to de-provision is rendered by Lotka-Volterra (Prey Decreasing Predator Increasing situation) algorithm. The VMs, which are not executing any cloudlets are selected and returned back to VM pool.
\subsection{Experiment}
The monitoring algorithm is implemented in Cloudsim 3.03 version. There is a class named DataCenterBroker, which is responsible for VM creation, cloudlets submission to a particular VM, destroying the VM once it executed all the cloudlets submitted to it, etc. The DataCenterBroker class is the perfect place, from where it is feasible to monitor the VMs utilization and addition and de-provisioning of VMs based on the pre-decided threshold. Few decisions are taken prior to the experiment that the cloudlets are going to be submitted dynamically. The VM pool number will be predefined and monitor will keep checking the VMs average utilization for every 100 milliseconds. There is a CloudletScheduler, which decides the request to be allocated to a VM. In this experiment , time shared Cloudlet scheduler is used to serve the purpose. This particular scheduler allocates a time slot for every submitted Cloudlet request. On the other hand, VmAllocationPolicySimple is responsible for the determination of the host where a VM will be created. A single data center is utilized through out the experiments. The data center consists of two host machines with one host machine powered by four processors(quad core), and the other host machine contains two processors(dual-core machine). The computation speed for each processor is 1000 MIPS. Hence the data center contains two host machines, where one host machine is quad core and another host machine is dual core. We have submitted cloudlets and VMs in three phases. First phase submits predefined cloudlets and VMs before starting the simulation. Other phases add more VMs and cloudlets intermittently after the beginning and before the completion of the simulation so that it can create a replica of a real time scenario. Apart form this, for every 100 milliseconds, 10 cloudlets are submitted during the simulation to maintain the continuation of incoming requests. DormandPrince853Integrator of the commons-math library is used to extract values from Lotka-Volterra model.
\begin{table}
\begin{center}
\begin{tabular}{c c c c c c} 
\hline 
Exp No &VM&cloudlets&Avg Req Compln time&SLA violation & MakeSpan time\\ 
\hline
 1&10 &98&514.1188& 0.744 &1697.94\\ 
1&15&135 & 686.6909&0.407&1446.94\\ 
1&16& 155 & 692.715&0.419&1657.25\\
2&10 &98&612.66& 0.744 &1648.94\\ 
2&15&135 & 336.72&0.29&1356.0\\ 
2&16& 165 & 333.785&0.315&1633.01\\ 
3&10 &98&550.51& 0.755 &1811.93\\ 
3&15&135 &334.84&0.37&1450.48\\ 
3&16& 165 & 312.19&0.32&1774.481\\
4&10 &98&611.77& 0.80 &1630.61\\ 
4&15&135 &317.35&0.31&1438.14\\ 
4&16& 165 & 315.72&0.30&1636.961\\
\hline
\end{tabular}
\caption{In the table, the performance of the Reactive scaling with LV modeling monitoring algorithm has been demonstrated. Three metrics Average Request Completion time, SLA violation rate and makespan time are displayed. Total three phases of submission are represented in the table for each experiment and above mentioned metrics are calculated for every phase. The experiments are conducted a total of 4 times and to maintain the uniformity, same VM, cloudlets numbers are used for every experiment. }\label{Exp1}
\end{center}
\end{table}
In the table \ref{Exp1}, three metrics, Average Request Completion time, SLA violation rate and makespan time are displayed. Total three phases of submission are represented in the table and above mentioned metrics are calculated for every phase.
Completion time is calculated as below
\begin{ceqn}
\begin{equation}\label{avgcompltntime}
Avg \quad Completion \quad time=\frac{\sum_{i=1}^{Cloudletnumber}Completion \quad time_{i}}{Number \quad of \quad Cloudlet}
\end{equation}
\end{ceqn}
For each phase the average completion time is calculated, whereas in equation \ref{avgcompltntime} number of cloudlets signifies the number of cloudlets of each phase. Makespan is the total duration between the beginning and the end. Here the makespan is defined as the time difference from the last request finish time and the first request submission time of a phase.

\begin{equation}\label{makespan}
makespan=finishing \quad time \quad of\quad last\quad request \\-submission\quad time\quad of\quad first\quad request
\end{equation}
Another parameter estimated along with makespan is SLA violation rate.
If the completion time of any request exceeds the SLA mentioned expected completion time, then the request violates the SLA. In such a scenario, the Cloud service provider has to pay for the SLA violation. Hence in an ideal scenario, SLA violation rate should be minimum. 
\begin{ceqn}
\begin{equation}\label{slaviorate}
SLA\quad violation\quad rate=\frac{number\quad of\quad requests\quad violates\quad SLA}{total\quad number\quad of\quad requests}
\end{equation}
\end{ceqn}




\subsection{VM Elasticity} \label{vmelasticity}
Elasticity is a term, which is very common in the field of Physics and Economics but now-a-days the same term is also frequently used in Cloud Computing. In the context of Cloud Computing, elasticity is defined as an ability to adapt to the changing workloads by provisioning and de-provisioning Cloud resources automatically such that it can meet the current demand of resources at any point of time \cite{Herbst}.
Elasticity is defined in physics as a property of material capturing the capability of returning to its original state after deformation.
In economics, elasticity refers to the sensitivity of a dependent variable to the one or more arguments \cite{CHIANG}. A brief and simple example will help to understand "How Elasticity plays its role in Cloud Computing". Consider a Website A, which is running in a certain data center and at that point of time $t_{0}$ as per the workload, 2 virtual machines are allocated. Due to the rising popularity of the website at time $t_{1}$, it has started receiving more requests and 2 virtual machines are not enough to serve all the requests. Hence, it needs to allocate more virtual machines. Consider now that 6 more virtual machines are required to cater to the changing workload.
An elastic system should identify the situation and allocate 6 virtual machines immediately. Several hours later, at $t_{3}$, the number of user requests dropped significantly and 3 virtual machines are sufficient to handle all the incoming user requests. In such a scenario, an elastic system should detect the change in incoming requests and de-provision 5 virtual machines allocated earlier. However, over-provisioning is a situation, where more resources are allocated than required. Such a situation needs to be avoided as service provider ends up spending more for the extra resources. It has dynamic impact on the optimal levels of service provision, because, as opposed to the previous case, under-provisioning is a situation where lesser number of resources are allocated for the service provider with serious impact on the quality of service. Indeed, it may often lead to violation of service level agreement and service provider loses jobs due to poor services.The economic losses faced is a direct outcome of this possibility. The scaling decision between time points must therefore attend to the allocation problem with better precision than is often the case.\\
\textbf{Reactive Scaling:} The methods used for resolving scaling decisions can be classified into two categories. One is Reactive methods and the other is Proactive methods. Rules (or threshold) define reactive resource allocation method used to determine limitations for violating a series of guidelines and measures related to resource scaling need to be carried out when these violations occur. Though, there are three major concerns with this method
\begin{itemize}
\item When rule violation happens, the scaling decision may involve SLA violation, which affects QoS.
\item It may also happen that scaling of decisions of resources due to some violations is not necessary, as the violations are temporary. Scaling up and down of resources is not required.
\item  The on-demand VM requires a certain amount of time to initialize, boot-up and start the applications. Therefore, new request for additional VM may fail as the VM was not ready within the required time-frame. 
\end{itemize}
In this experiment, the monitoring algorithm analyzes the VMs utilization for every equal interval of time. In case it encounters a violation, it can take the right step to mitigate the situation. If the average utilization violates the maximum threshold then it is going to add a VM from the VMs pool, whereas the violation of minimum threshold will eradicate a VM and return it to the VM pool. As indicated in the tables, the same number of sets are used across, and the experiment has been repeated four times, which has produced results with small variations. Like Lotka-Volterra monitoring algorithm, the same metrics (Average Request Completion time, SLA violation, Makespan time) are explored in Reactive scaling algorithm.  

\textbf{Resource Optimality:} Now, if we compare the result of Lotka-Volterra with Reactive scaling algorithm, it is evident from the tables that except for a few instances, Lotka-Volterra algorithm has outperformed the Reactive scaling algorithm with respect to all the three metrics. \cite{Aslanpour} has shown that the average utilization of VM is 55.1\% during reactive scaling. The configuration of the experiment is as follow:
MaxThreshold is set to 80\% while MinThreshold is set to 20\%, MaxVMallocated= 18 and Monitoring interval is 1 minute. It has been concluded that scaling rule based on CPU utilization caused poor performance due to high number of scaling decisions. In contrast, ALVEC produces 95\% VM utilization in the case of reactive scaling and much lower scaling decisions, which in turn improves the performance of reactive scaling algorithm. The similar threshold values are employed thoughout the simulation which lasts for few milliseconds and involves higher number of VM's. 

\begin{table}
\begin{center}
\begin{tabular}{c c c c c c} 
\hline 
Exp No&VM&cloudlets&Avg Req Compln time&SLA violation & MakeSpan time\\ 
\hline
 1&10 &98&814.08& 0.89 &1902.4\\ 
1&15&135 &339.068&0.34&1432.05\\ 
1&16& 155 & 466.15&0.49&1902.91\\ 
2&10 &98&686.63& 0.82 &1726.91\\ 
2&15&135 &371.87&0.41&1591.05\\ 
2&16& 155 & 384.457&0.37&1724.37\\
3& 10 &98&788.20& 0.89 &1826.92\\ 
3&15&135 &297.29&0.207&1315.98\\ 
3&16& 155 & 400.55&0.361&1811.34\\
4&10 &98&786.21&0.877 &1867.93\\ 
4&15&135 &341.06&0.325&1509.96\\ 
4&16& 155 &435.70&0.451&1861.81\\
\hline
\end{tabular}
\caption{The table showcases the performance of the Reactive scaling algorithm without LV model. Total four experiments are displayed and each experiment comprises of three phases. Three metrics: average request completion time, SLA violation rate and makespan are evaluated for every experiment.  }\label{reactive1}
\end{center}
\end{table}

\begin{figure}[!tbp]
  \centering
  \begin{minipage}[b]{0.45\textwidth}
    \includegraphics[width=\textwidth]{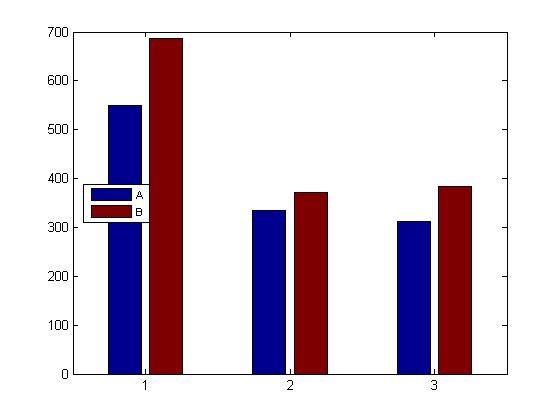}
\caption{The Y axis represents the average completion time. 1, 2 and 3 of X axis denotes the first, second and third phase respectively. A in the figure depicts the Reactive scaling with Lotka-Volterra average completion time and B signifies the reactive scaling without LV average completion time.
It is visible that A has performed better than B in all the three phases.
The dataset of two tables, exp no 2 of table \ref{Exp1} and exp no 2 of \ref{reactive1} are used in the figure, which proves the superiority of Lotka-Volterra over the other algorithm.}
\label{monitoringavg}
\centering
  \end{minipage}
  \hfill
  \begin{minipage}[b]{0.45\textwidth}
    \includegraphics[width=\textwidth]{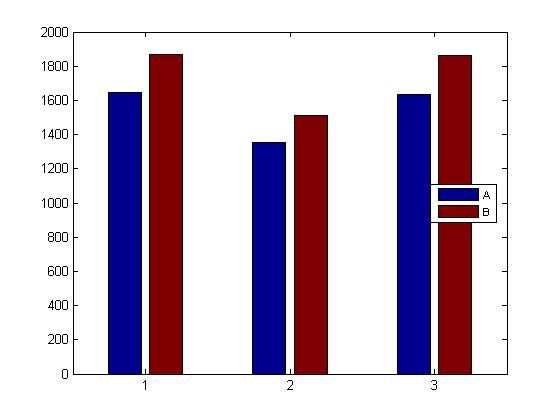}
\caption{depicts the makespan comparison of the two algorithms. Like the previous figure, A and B are representing Lotka-Volterra and Reactive scaling respectively. The makespan time is shown alongside the Y axis whereas the X axis shows the three phases. The first phase consists of 98 cloudlets. 135 cloudlets are part of second phase whereas third phase consists of 165 cloudlets. A has outperformed B in makespan duration as the makespan duration of A is lesser than B. Table \ref{Exp1}, exp no 2 and Table \ref{reactive1}, exp no 4 are illustrated in the makespan comparison figure.}
\label{monitormakespan}
\centering
  \end{minipage}
\end{figure}
\begin{figure}[h!]
\centering
\includegraphics[width=0.5\textwidth]{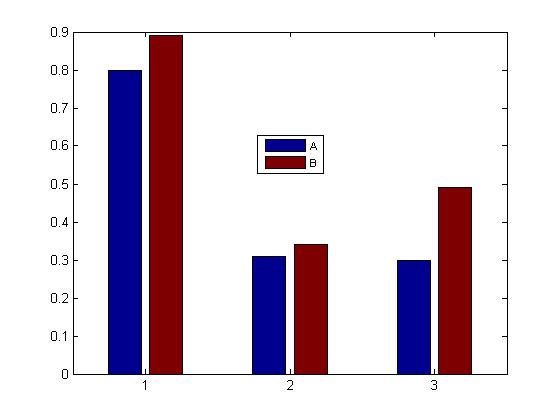}
\caption{Illustrates the SLA violation comparison between the two algorithms. Table \ref{Exp1}, exp no 4 and Table \ref{reactive1}, exp no 4 are discussed in the SLA comparison figure. Deadline considered for each phase is 400 milliseconds. A request would violate the SLA, if its completion time exceeds the predefined time, which is 400 milliseconds here. Each phase SLA violation numbers of both the algorithms have been displayed side by side. A depicts the Lotka-Volterra algorithm and B signifies Reactive scaling algorithm. It is evident from the figure that SLA violation rate is lesser in the case of A than in B.}
\label{monitorsla}
\centering
\end{figure}

\subsection{Proactive Scaling}\label{responsemakespan}
In this section, we are going to discuss and compare the performance of Proactive scaling after integration with Lotka-Volterra model. Unlike reactive scaling, the QoS parameter that has been considered for threshold calculation is response time. In this experiment, the monitoring algorithm predicts the response time in future for equal time intervals. If the monitoring algorithm identifies any SLA violation of response time, an upscaling/downscaling decision has to be made. Response time threshold is predefined and decided when the SLA is made between Cloud provider and user. Proactive scaling requires two types of thresholds. One is upper threshold and the other is lower threshold. If any situation arises where upper threshold is violated by future response time prediction, new VMs have to be added to service to neutralize the situation. In the case of lower threshold violation, VMs need to be deallocated from the user service as more than required number of VMs are allocated to increase utilization of the resources. We have compared the performances between the two cases. In one case, Lotka-Volterra model is employed to calculate the number of VMs that need to be added or removed based on threshold violation. In any case, a VM is allocated/ deallocated based on SLA violation prediction. The method, which is used for the prediction of future response time is WMA (Weighted Moving Average). It is widely used and very familiar in stock market strategy. It has multiplying factors to give different weights to data at different positions \cite{Oriol}.
\begin{ceqn}
\begin{equation}\label{wma}
WMA(t)=\frac{n*data_{t-1}+(n-1)*data_{t-2}+..+2*data_{t-n+2}+data_{t-n+1}}{n+(n-1)+(n-2)+..+2+1}
\end{equation}
\end{ceqn}
 
\begin{table}
\begin{center}
\begin{tabular}{c c c c c c c} 
\hline 
\thead{No}&\thead{SLA VLTN\\ WO LV}&\thead{SLA VLTN\\ W LV}&\thead{ExETN time \\WO LV} & \thead{ExETN time\\ W LV}& \thead{Makespan WO\\ LV}&\thead{Makespan\\ W LV}\\ 


\hline
1st Scenario &.458&.43& 1711.95 &1572.10&3046.74&3024.26\\ 
2nd Scenario&.535 &.45&529.72&453.81&2838.86&2271.20\\ 
\hline

\end{tabular}
\caption{Lotka-Volterra Proactive algorithm comparison with Proactive scaling without LV has been illustrated in table \ref{Proactive}. SLA VLTN means SLA Violation, ExETN time stands for Execution time, WO LV represents without LV model and W LV denotes with LV model. Total 630 cloudlets are used during 1st scenario simulation, where 550 cloudlets are fed into the data center into three phases. 1st phase has pushed 98 cloudlets, 2nd phase consists of 302 cloudlets and third phase inserted 150 cloudlets. Rest of the 80 cloudlets are generated during simulation time. The SLA deadline for all the cloudlets execution time, has been decided as 2000 milliseconds. The total number of VM's allocated initially is 27 and 100 VMs in pool, which will be used in elasticity algorithm for further allocation based on situation. Upper threshold for scaling decision has been set at 400 ms and lower threshold is 100 ms. 2nd scenario consists of total 394 cloudlets as an input to the data center, where all the cloudlets are  pushed into three batches. First batch consists of 12 cloudlets, second batch consists of 102 cloudlets and third batch pushed 200 cloudlets. 80 cloudlets are generated during simulation time. Total number of VM in VM pool is 100. SLA violation has been set at 500 milliseconds for execution time quality of parameter. Like scenario 1, 27 VMs are allocated initially. Though the cloudlets are part of three batches, all the cloudlets across the batches are submitted to data center broker for execution at random time, which makes the entire process interactive similar to the real time data center operation. Both the scenarios are applied on predictive scaling with and without Lotka-Volterra model. It is evident from the table that proactive scaling with LV has outperformed its counter part in all three phases.}\label{Proactive}
\end{center}
\end{table}

\begin{figure}[!tbp]
  \centering
  \begin{minipage}[b]{0.45\textwidth}
    \includegraphics[width=\textwidth]{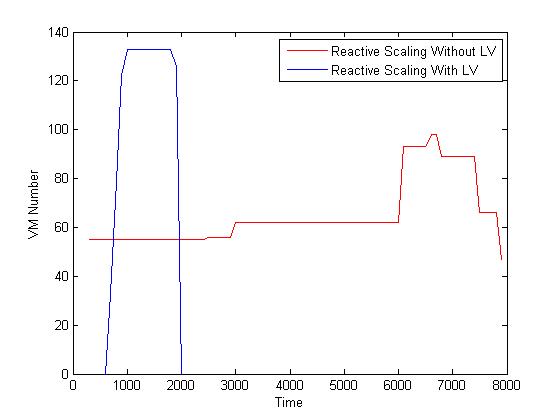}
\caption{This figure depicts the comparison of reactive scaling algorithm with and without LV model. Y axis represents the VM number and X axis represents the time. It is apparent from the figure that the simulation of  reactive algorithm with LV lasts for shortest period of time, till 2000 milliseconds. On the other hand, the reactive algorithm without LV goes on till 8000 milliseconds. But LV model uses maximum number of VMs to complete all the cloudlets task.}
\label{ReactiveLkVsWithoutLk}
\centering
  \end{minipage}
  \hfill
  \begin{minipage}[b]{0.45\textwidth}
    \includegraphics[width=\textwidth]{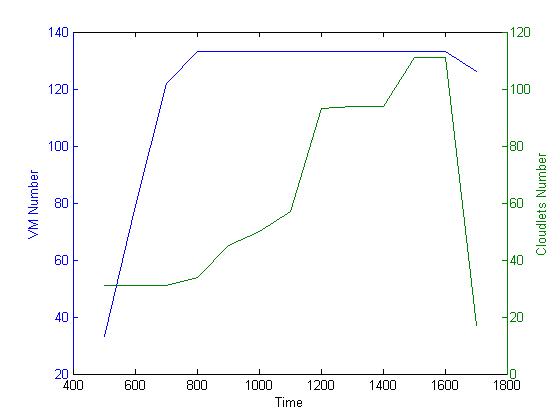}
\caption{The figure illustrates the VM number and Cloudlet number( represents jobs) against time for reactive scaling algorithm with LV model. X axis plots the Time and Y axis left side and right side represent the VM number and Cloudlet number respectively. The figure demonstrates the behavior of the VM and cloudlets as simulation progresses. The blue curve denotes the VM number whereas the green curve showcases the Cloudlet number. Initially, both the curves rise but VM number reaches it's threshold limit, hence the curve becomes flat afterwards. But VM number curve starts declining as the Cloudlet number starts falling.}
\label{VMvsCloudletVsTime}
\centering
  \end{minipage}
\end{figure}

\begin{figure}[h!]
\centering
\includegraphics[width=0.6\textwidth]{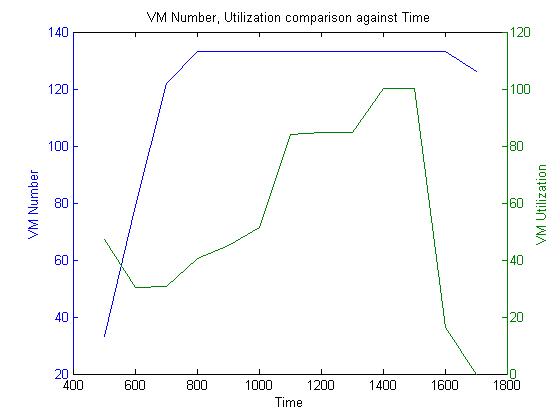}
\caption{The figure depicts the VM number, VM utilization against time \textbf{by using Algorithm \ref{paramtertuning}}. X axis denotes the time. Y axis left side represents the VM number and Y axis right side represents the VM utilization. In the first phase the utilization starts falling as more VMs are being allocated due to the rising number of cloudlets(jobs). But the VM utilization rises up as many more cloudlets start arriving and VM number reaches its threshold level. At the last lag, both the curves(VM number, VM utilization) start declining as cloudlets arriving rate reduces. Blue, green curve depict the VM number, VM utilization respectively.}
\label{VmVsUtilizationVsTime}
\centering
\end{figure}
\subsection{Predator-Prey cloudlets(job) Scheduling Timeshared Algorithm}
In this section\footnote{Predtor is jobs i.e cloudlets, prey is resource i.e. VM}, a Cloudlet time sharing algorithm is explored in a new dimension. We have integrated the Lotka-Volterra model with the existing Cloudlet time sharing algorithm of Cloudsim and have taken advantage of the predator-prey equation. The algorithm decides the VM occupancy before submitting the incoming Cloudlet request to VM. If the occupancy or the VM utilization is more than the predefined threshold, Lotka-Volterra (Prey increasing-Predator deceasing) model is invoked to retrieve  VM number, which is required to reduce the utilization within threshold.
Here, prey denotes the number of available VM and predator the number of cloudlets. Till the time the VM number has not reached the Lotka-Volterra suggested number, all the incoming requests are pushed into a waiting list to the corresponding VM. Once the normal situation attains, that is the utilization drops below threshold, the requests from waiting list are sent for execution. In case the utilization of VM goes down the minimum threshold and incoming requests are available then more cloudlets requests are submitted to VMs for processing as per Lotka-volterra model(Prey decreasing-Predator increasing situation). This particular situation is applicable when incoming requests are available. The formula used to identify the VMs utilization is below:
\begin{ceqn}
\begin{equation}\label{vmutil}
\begin{split}
vm\quad utilization=  \frac{total\quad vm\quad executing\quad the\quad requests\quad at\quad that\quad moment}{total\quad number\quad of\quad available \quad VMs }
\end{split}
\end{equation}
\end{ceqn}
\begin{algorithm}
\caption{LV-Timeshared Algorithm}\label{alg:lv-timeshared}
\begin{algorithmic}[1]
\Procedure{LV-Timeshared}{}\Comment{}
\State $scloudlets \gets cloudlets-submitted-for-execution $
\State $ cloudlets_Q \gets existing-cloudlets-in-queuee $
\State $ T \gets Time-Allocated $
\State $oVM \gets occupancy-of-VMs $

\While{$ T \neq 0 $}

\State call procedure ${LV-PARAMETER-TUNING} $ 
 
 \If{$oVMs > MaxThreshold$} \\ \Comment{VM increasing, cloudlets decreasing}
 
 \While{$ oVM > available-VM $}
 \State add $ cloudlets_Q \gets scloudlets +cloudlets_Q $
 \EndWhile 
 
\EndIf

\If{$oVMs < MinThreshold$} \\ \Comment{VM decreasing, cloudlets increasing}
  \State add $ cloudlets_Q \gets scloudlets +cloudlets_Q $
 \EndIf
\State $ T \gets  T-1 $

\EndWhile
\State \textbf{return} 
\EndProcedure
\end{algorithmic}
\end{algorithm}

\underline{\textbf{Predator-Prey cloudlets Scheduling Timeshared Algorithm}}\footnote{Predtor is jobs i.e cloudlets, prey is resource i.e. VM}
\begin{itemize}
\item cloudlets submitted for execution.
\item Calculate occupancy of VMs.
\item If(occupancy of $VMs>thresholds$) [very less number of VMs available]
\item call Lotka-Volterra for Prey Increasing-Predator Decreasing [VM number will increase and Cloudlet will decrease]
\item add cloudlets to waiting queue till it drops to LV suggested VM occupancy/surges till LV suggested available VMs.
\item Once the LV suggested VM number reaches, start submitting the  cloudlets  from waiting queue.
\item If the occupancy drops below the required level(Many VMS are available)
\item Call Lotka-Volterra for Prey decreasing-Predator Increasing (VM occupancy will increase/available VM will reduce and Cloudlet will increase)
\item Submit cloudlets without keeping them in waiting queue till it ramps up to the LV suggested number.(This situation will work in case incoming cloudlets are available)
\end{itemize} 

\subsection{Simulation of Timeshared Algorithm}
Two metrics have been considered for the experiment purpose. These are average Cloudlet completion time and SLA violation rate. The Lotka-Volterra time shared algorithm has been compared with the existing Cloudlet time sharing algorithm of Cloudsim and it is shown how performance of the algorithm is improved. The cloudlets are part of three batches but all the cloudlets are submitted to the broker of data center at random time. This mimics the real time scenario, where we don't have any control on the arrival time of cloudlets.
\par \textbf{Experiment 1:}\\
In this experiment, the VM and cloudlets are submitted in three batches. The table explains all the batches and the corresponding VM and cloudlets numbers of each batch. Deadline is a predefined number for each Cloudlet request before the execution starts. If the Cloudlet execution time surpasses the deadline, it is considered as SLA violation. For 1st, 2nd and 3rd  batch, the deadlines are 450, 1000 and 1000 milliseconds.This experiment has been demonstrated in table \ref{Exp1Timeshared}.
\begin{table}[!tbp]
\begin{center}
\begin{tabular}{c c c c c c c} 
\hline 

\thead{VM} & \thead{Cloudlets} & \thead{LV Time\\ Sharing  Avg.\\ Execution Time} & \thead{Cloudlet Time \\Sharing  Scheduling \\ avg execution time} & \thead{Deadline} & \thead{LV Cloudlet \\ time sharing \\ SLA violation} & \thead{Cloudlet Time\\ Sharing  SLA \\violation} \\
\hline
60 &80&437.97& 472.84 &450&0.2&0.3\\ 
55&180 & 873.007&1069.92&1000&0.37&0.65\\ 
9& 87 & 876.26&1066.63&1000&0.45&.70\\ 
\hline
\end{tabular}
\caption{This table comprises of three batch executions of VM and cloudlets. Each row represents one batch execution. The comparison between Time shared scheduling algorithm of Cloudsim and Time shared scheduling algorithm with LV has been demonstrated. Two metrics, SLA violation rate and avg execution time are highlighted in the table. It is evident from the table data that time shared scheduling algorithm with LV has outperformed the other time shared algorithm. }\label{Exp1Timeshared}
\end{center}
\end{table}

\begin{figure}[!tbp]
  \centering
  \begin{minipage}[b]{0.45\textwidth}
    \includegraphics[width=\textwidth]{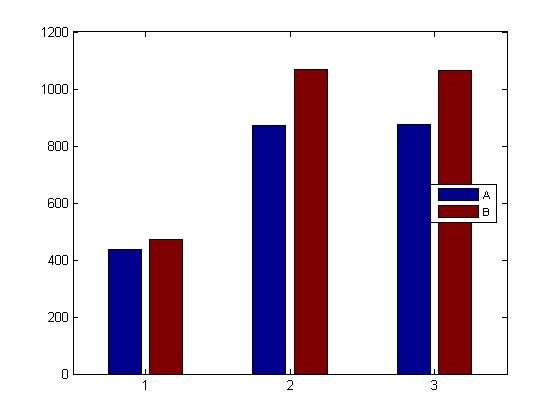}
\caption{Average Completion time Comparison between time shared scheduling algorithm with LV and without LV. A denotes the time shared scheduling algorithm with LV and B represents the time shared scheduling algorithm without LV. For all the batches, the average completion time is better for time shared LV algorithm. Y axis represents the average completion time and X axis the batches.}
\label{Cloudschedulertime1}
\centering
  \end{minipage}
  \hfill
  \begin{minipage}[b]{0.45\textwidth}
    \includegraphics[width=\textwidth]{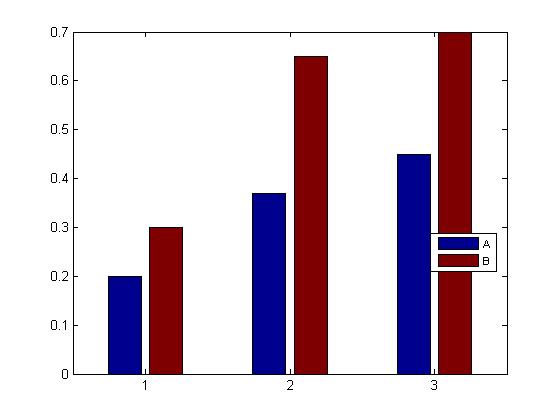}
\caption{SLA Comparison between time shared scheduling algorithm with LV and without LV. A, B represent time shared scheduling algorithm with LV and time shared scheduling algorithm without LV, respectively. SLA violation rate is less in the case of LV time shared algorithm.}
\label{Cloudschedulersla1}
\centering
  \end{minipage}
\end{figure}
\par \textbf{Experiment 2}\\
In the second experiment, VM and Cloudlet population for each batch have changed. The deadlines for execution time are set at 450, 3000, 3000 milliseconds for 1st, 2nd and 3rd batch. Table \ref{Exp2Timeshared} shows the result of experiment 2.
\begin{table}[!tbp]
\begin{center}
\begin{tabular}{c c c c c c c} 
\hline 
\thead{VM} & \thead{Cloudlets} & \thead{LV Time\\ Sharing  Avg.\\ Execution Time} & \thead{Cloudlet Time \\Sharing  Scheduling \\ avg execution time} & \thead{Deadline} & \thead{LV Cloudlet \\ time sharing \\ SLA violation} & \thead{Cloudlet Time\\ Sharing  SLA \\violation} \\
\hline
 60 &80&459.82& 478.45 &450&0.225&0.275\\ 
55&300 & 2229.98&2654.66&3000&0.096&0.24\\ 
9& 245 & 2300.13&2659.14&3000&0.11&.15\\ 
\hline
\end{tabular}
\caption{This table comprises of three batch executions of VM, cloudlets. Each row represents one batch execution. The comparison between Time shared scheduling algorithm of Cloudsim and Time shared scheduling algorithm with LV has been demonstrated. Two metrics SLA violation rate and average execution time are highlighted in the table. It is evident from the table data that time shared scheduling algorithm with LV has outperformed the other time shared algorithm.  }\label{Exp2Timeshared}
\end{center}
\end{table}
\begin{figure}[!tbp]
  \centering
  \begin{minipage}[b]{0.45\textwidth}
    \includegraphics[width=\textwidth]{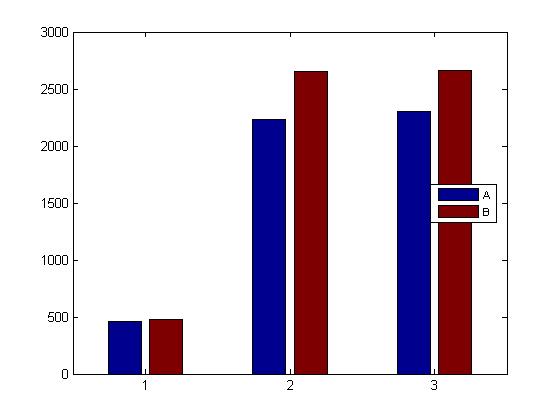}
\caption{Average Completion time Comparison between time shared scheduling algorithm with LV and without LV. A denotes the time shared scheduling algorithm with LV and B represents the time shared scheduling algorithm without LV. For all the batches, the average completion time is better for time shared LV algorithm. Y axis represents the average completion time and X axis the batches. }
\label{Cloudschedulertime2}
\centering
  \end{minipage}
  \hfill
  \begin{minipage}[b]{0.45\textwidth}
    \includegraphics[width=\textwidth]{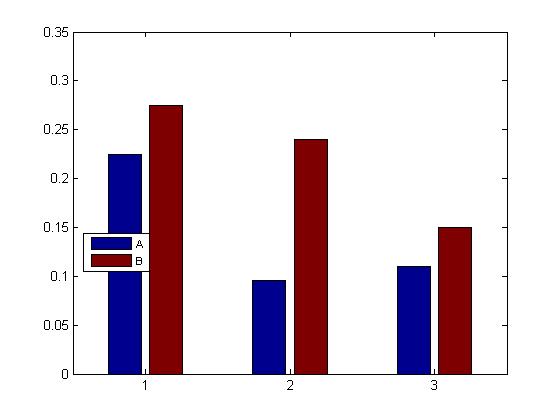}
\caption{SLA Comparison between time shared scheduling algorithm with LV and without LV. A, B represent time shared scheduling algorithm with LV and time shared scheduling algorithm without LV. SLA violation rate is less in the case of LV time shared algorithm.}
\label{Cloudschedulersla2}
\centering
  \end{minipage}
\end{figure}

\textbf{Experiment 3}\\
In this experiment, different VM and Cloudlet population are considered for second and third batch, though VM's and cloudlets from the first batch are uniform across the experiments. Deadline for each batch is similar to the experiment 1. \textbf{The outcome of the experiment has been displayed in table \ref{Exp3Timeshared}}.

\begin{table}[!tbp]
\begin{center}
\begin{tabular}{c c c c c c c} 
\hline 
\thead{VM} & \thead{Cloudlets} & \thead{LV Time\\ Sharing  Avg.\\ Execution Time} & \thead{Cloudlet Time \\Sharing  Scheduling \\ avg execution time} & \thead{Deadline} & \thead{LV Cloudlet \\ time sharing \\ SLA violation} & \thead{Cloudlet Time\\ Sharing  SLA \\violation} \\
\hline
 60 &80&462.35& 463.43 &450&0.375&0.325\\ 
55&90 & 928.47&1066.84&1000&0.5&0.6\\ 
9& 175 & 925.68&1121.98&1000&0.53&0.662\\ 
\hline
\end{tabular}
\caption{This table comprises of three batch execution of VM and cloudlets. Each row represents one batch execution. The comparison between Time shared scheduling algorithm of Cloudsim and Time shared scheduling algorithm with LV has been demonstrated. Two metrics SLA violation rate and average execution time are highlighted in the table. It is evident from the table data that time shared scheduling algorithm with LV has outperformed the other time shared algorithm.  }\label{Exp3Timeshared}
\end{center}
\end{table}

\begin{figure}[!tbp]
  \centering
  \begin{minipage}[b]{0.45\textwidth}
    \includegraphics[width=\textwidth]{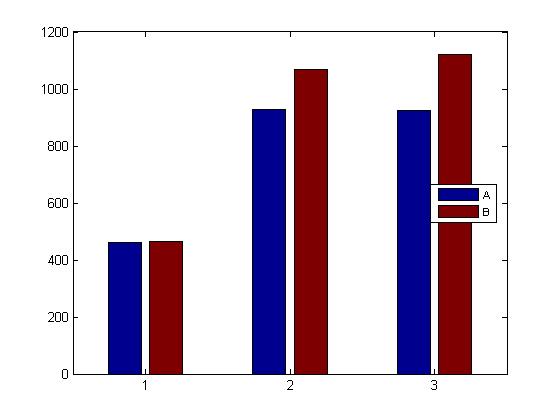}
\caption{Average Completion time Comparison between time shared scheduling algorithm with LV and without LV. A denotes the time shared scheduling algorithm with LV and B represents the time shared scheduling algorithm without LV. For all the batches, the average completion time is better for time shared LV algorithm. Y axis represents the average completion time and X axis the batches.}
\label{Cloudschedulertime3}
\centering
  \end{minipage}
  \hfill
  \begin{minipage}[b]{0.45\textwidth}
   \includegraphics[width=\textwidth]{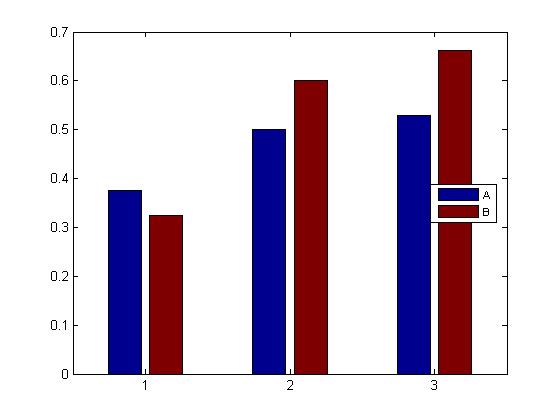}
\caption{SLA Comparison between time shared scheduling algorithm with LV and without LV. A, B represent time shared scheduling algorithm with LV and time shared scheduling algorithm without LV, respectively. SLA violation rate is improved after introducing LV model into time shared scheduling algorithm. }
\label{Cloudschedulersla3}
\centering
  \end{minipage}
\end{figure}
\begin{figure}[h!]

\end{figure}

\begin{figure}[h!]

\end{figure}

\begin{figure}[!tbp]
  \centering
  \begin{minipage}[b]{0.45\textwidth}
    \includegraphics[width=\textwidth]{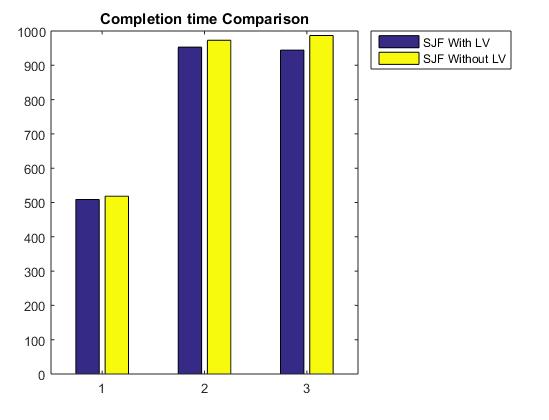}
\caption{\textbf{The comparisons between  SJF with LV ad without LV have been demonstrated. Each column represents one batch. Introduction of LV has improved the avg. completion time of SJF algorithm.}}
\label{completioSJF}
\centering
  \end{minipage}
  \hfill
  \begin{minipage}[b]{0.45\textwidth}
   \includegraphics[width=\textwidth]{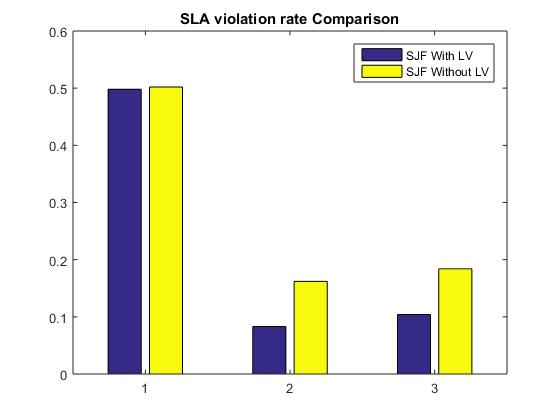}
\caption{ \textbf{The SLA violation rate has been used as a performance measurement metric in the above graph. SJF with LV has shown significant improvement in SLA violation rate.} }
\label{slaSJF}
\centering
  \end{minipage}
\end{figure}
\begin{figure}[h!]

\end{figure}

\begin{figure}[!tbp]
  \centering
  \begin{minipage}[b]{0.45\textwidth}
    \includegraphics[width=\textwidth]{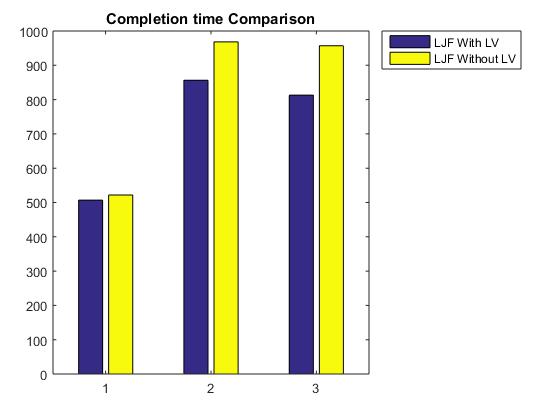}
\caption{The comparisons of LJF with LK and without LK have bee displayed in the above figure. Y axis represents the avg. completion time and the X axis represents the batch number. The performance of LJF with LK is better than LJF without LK.}
\label{completionLJF}
\centering
  \end{minipage}
  \hfill
  \begin{minipage}[b]{0.45\textwidth}
   \includegraphics[width=\textwidth]{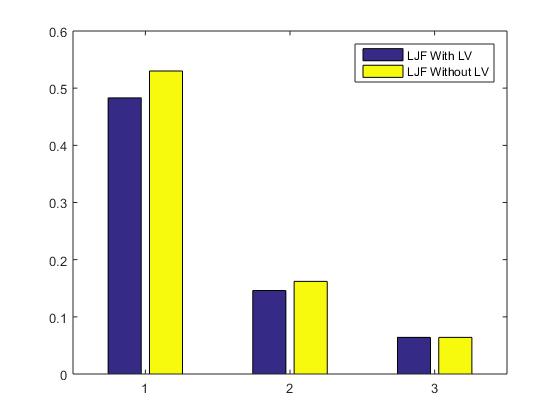}
\caption{The improvement of SLA violation rate of LJF algorithm after the introduction of LV has been showcased in the above figure.}
\label{slaLJF}
\centering
  \end{minipage}
\end{figure}
\begin{figure}[h!]

\end{figure}
\begin{figure}[!tbp]
  \centering
  \begin{minipage}[b]{0.45\textwidth}
    \includegraphics[width=\textwidth]{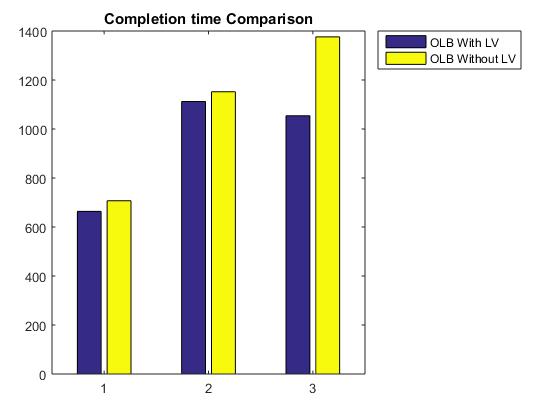}
\caption{\textbf{The  avg. completion time of both the algorithms, OLB with LV and without LV has been demonstrated in the above graph. The avg. completion time has been reduced in case of OLB with LV. The X, Y axises are representing the batch number and avg. completion time.}}
\label{completionOLB}
\centering
  \end{minipage}
  \hfill
  \begin{minipage}[b]{0.45\textwidth}
   \includegraphics[width=\textwidth]{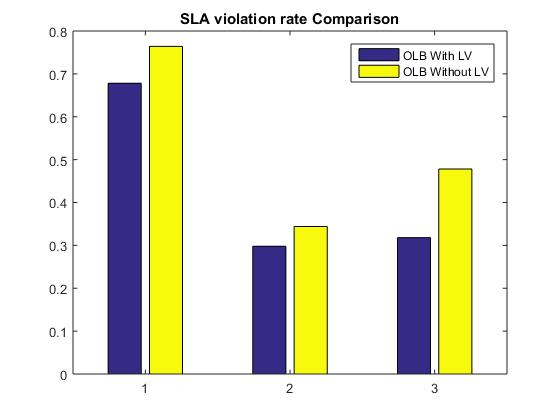}
\caption{ \textbf{X axis signifies the batch number whereas the Y axis signifies the SLA violation rate. The improvement is visible in the graph due to the introduction of LV.}}
\label{slaOLB}
\centering
  \end{minipage}
\end{figure}
\begin{figure}[h!]

\end{figure}

\begin{figure}[!tbp]
  \centering
  \begin{minipage}[b]{0.45\textwidth}
    \includegraphics[width=\textwidth]{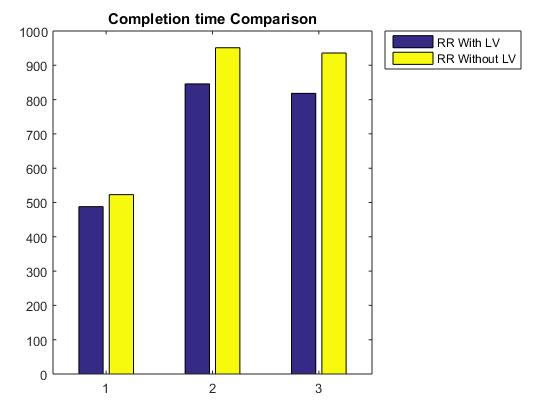}
\caption{\textbf{The avg. completion time reduction has been observed after the induction of LV into RR algorithm. Like other graphs, the X, Y axises represent the batch number and avg. completion time.} }
\label{completionTimeRR}
\centering
  \end{minipage}
  \hfill
  \begin{minipage}[b]{0.45\textwidth}
   \includegraphics[width=\textwidth]{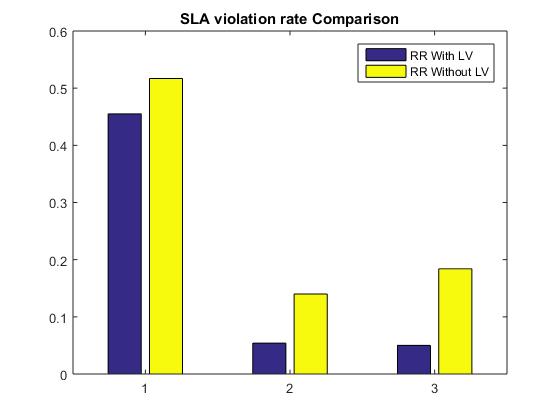}
\caption{\textbf{Like avg. completion time, the SLA violation rate improvement is evident from the figure, where the Y axis stands for SLA violation rate and X axis represents the batch number.}}
\label{slaRR}
\centering
  \end{minipage}
\end{figure}
\begin{figure}[h!]

\end{figure}

\begin{figure}[!tbp]
  \centering
  \begin{minipage}[b]{0.45\textwidth}
    \includegraphics[width=\textwidth]{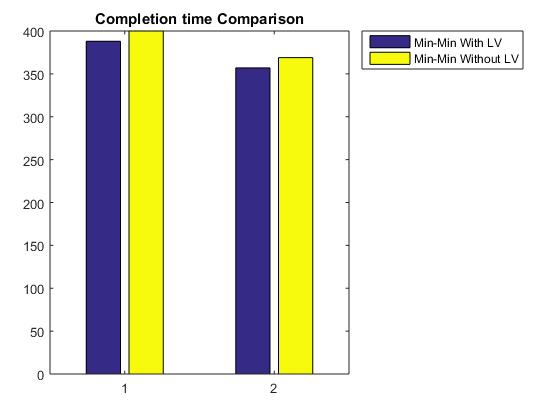}
\caption{The graph illustrates the Min-mi algorithm avg. completion time performance comparison after the adoption of LV. The first column represents task completion time, where the MIPs of the VMs were not fixed or predefined rather dynamic during simulation. The second column demonstrate the avg. completion time where the MIPs of VMs were predetermined.In both the instances, LV has improved the performance of Min-min algorithm}
\label{completionTimeMinMin}
\centering
  \end{minipage}
  \hfill
  \begin{minipage}[b]{0.45\textwidth}
   \includegraphics[width=\textwidth]{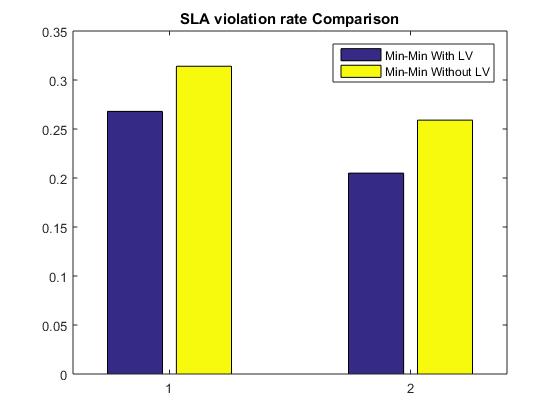}
\caption{Another QoS metric, SLA  violation rate of MIn-min algorithm has been displayed in the above figure, where in the first column MIPs of the VMs were not fixed and in the second column the MIPs of the VMs were predetermined. }
\label{slaMinMin}
\centering
  \end{minipage}
\end{figure}
\begin{figure}[h!]

\end{figure}

\begin{table}[!tbp]
\begin{center}
\begin{tabular}{c c c c c c c} 
\hline 
\thead{VM} & \thead{Cloudlets} & \thead{SJF with LV \\   Avg.\\ Execution Time} & \thead{SJF without LV\\ avg execution time} & \thead{Deadline} & \thead{SJF with LV \\ SLA violation} & \thead{SJF without LV\\  SLA \\violation} \\
\hline
 60 &80&508.73& 516.4 &450&0.49&0.50\\ 
5&35 & 952.89&973&1000&0.08&0.16\\ 
9& 15 & 944.42&987&1000&0.10&0.18\\ 
\hline
\end{tabular}
\caption{The table exhibits the performance of the SJF algorithm in terms of QoS metrics such as SLA violation rate, avg. completion time.The performance of the SJF algorithm has improved significantly after the introduction of LV in the afore mentioned algorithm.}\label{sjftable}
\end{center}
\end{table}

\begin{table}[!tbp]
\begin{center}
\begin{tabular}{c c c c c c c} 
\hline 
\thead{VM} & \thead{Cloudlets} & \thead{LJF with LV \\   Avg.\\ Execution Time} & \thead{LJF without LV\\ avg execution time} & \thead{Deadline} & \thead{LJF with LV \\ SLA violation} & \thead{LJF without LV\\  SLA \\violation} \\
\hline
 60 &80&507.8& 522.6 &450&0.483&0.53\\ 
5&35 & 856.4&968.2&1000&0.146&0.162\\ 
9& 15 & 813&957.2&1000&0.064&0.064\\ 
\hline
\end{tabular}
\caption{The performance comparisons of LJF algorithm  are displayed in the above table. It is prominent from the table that the LJF algorithm's performance improved due to the induction of LV.}\label{ljftable}
\end{center}
\end{table}

\begin{table}[!tbp]
\begin{center}
\begin{tabular}{c c c c c c c} 
\hline 
\thead{VM} & \thead{Cloudlets} & \thead{OLB with LV \\   Avg.\\ Execution Time} & \thead{OLB without LV\\ avg execution time} & \thead{Deadline} & \thead{OLB with LV \\ SLA violation} & \thead{OLB without LV\\  SLA \\violation} \\
\hline
 60 &80&664& 707 &450&0.678&0.764\\ 
5&35 & 1112.2&1152.6&1000&0.298&0.344\\ 
9& 15 & 1054.6&1376.2&1000&0.318&0.478\\ 
\hline
\end{tabular}
\caption{ The table comprises of three rows and each row represents one batch execution. Therefore total three executions are showcased. It is evident from the table that the performance of OLB with LV better than OLB without LV.}\label{olbtable}
\end{center}
\end{table}

\begin{table}[!tbp]
\begin{center}
\begin{tabular}{c c c c c c c} 
\hline 
\thead{VM} & \thead{Cloudlets} & \thead{RR with LV \\   Avg.\\ Execution Time} & \thead{RR without LV\\ avg execution time} & \thead{Deadline} & \thead{RR with LV \\ SLA violation} & \thead{RR without LV\\  SLA \\violation} \\
\hline
 60 &80&504& 523 &450&0.475&0.517\\ 
5&35 & 846&951&1000&0.05&0.14\\ 
9& 15 & 818.2&936.6&1000&0.05&0.184\\ 
\hline
\end{tabular}
\caption{RR algorithm performance has been displayed in the above table. Each row exhibits one batch execution. Total two QoS metrics, avg. completion time, SLA violation rate are evaluated after the induction of LV in the RR algorithm.}\label{rrtable}
\end{center}
\end{table}

\begin{table}[!tbp]
\begin{center}
\begin{tabular}{c c c c c c c} 
\hline 
\thead{VM} & \thead{Cloudlets} & \thead{Min-Min with LV \\   Avg.\\ Execution Time} & \thead{Min-min\\ without LV\\ avg execution time} & \thead{Deadline} & \thead{Min-min\\ with LV \\ SLA violation} & \thead{Min-min \\without LV  SLA \\violation} \\
\hline
 60 &120&357& 369 &450&0.2052&0.517\\
 60 & 120&388.2&400.8&450&.268&.3142\\
 
\hline
\end{tabular}
\caption{The table comprises of two rows, each row demonstrates a different situation. First row shows the instance where the MIPs of the VMS are predetermined or predefined before the simulation. In the second row highlights the instance where the MIPs of the VMs are not fixed. Two QoS metrics, avg. completion time, SLA violation rate are compared in the table.   }\label{min-mintable}
\end{center}
\end{table}


\subsection{Improved Quality of Service \& Reduction in SLA Violation}
\begin{itemize}
\item Elasticity: It is an ability of a a Cloud data center to provision and de-provision VM as per the dynamic behavior of the Cloud resource demand. Lotka-Volterra provides the flexibility to decide the required number of VMs needed to be introduced into the systems based on the number of current Cloud resource requests. VM Elasticity is discussed in detail in "Model Implementation and Outcome: Technical Discussion" section.
\item Make span : Makespan is an another metric, used in this paper to measure the performance improvement of various algorithms such as reactive scaling, proactive scaling, Cloudlet time shared algorithm, etc, after the introduction of the Lotka-Volterra model into the afore mentioned algorithms. Reference of makespan can be observed across the paper such as subsection VM Elasticity, Proactive Scaling under "Model Implementation and Outcome: Technical Discussion" section.
\item Response time : It is a widely used QoS parameter in Cloud computing. We have shown that the implementation of Lotka-Volterra model has improved the response time performance metric significantly. Multiple occurrences of this particular QoS can be observed in various sections of this paper such as "Model Implementation and Outcome: Technical Discussion".
\item Utilization : Utilization has been used heavily in this paper. In the algorithms such as LV-Timeshared algorithm, the invocation of Lotka-Volterra has occurred whenever the utilization is touching the maximum/ minimum predefined threshold. 
\item Reduction in SLA violation: SLA violation is the most important performance metric of a Cloud data center. Revenue of a Cloud service provider is tightly coupled with SLA violation rate. After the introduction of LV model into various algorithms, SLA violation rate has been reduced, which has improved the quality of service and job satisfaction. More details can be found in "Model Implementation and Outcome: Technical Discussion" section.
\end{itemize}

\subsection{Predictive Analytics on VM population-LS fitting}
Predictive analysis of the requirements of VM's based on the demand of jobs is a handy tool to have. However, since the ALVEC model is nonlineaar in nature, simulated data poses some challenges when fitting algorithms are attempted. Standard algorithms didn't work. However, the scatter plot of the simulated data (generated due to the interaction between VM and jobs) is enlightening. The dataset consists of two groups of points which correspond to elliptical curves. A good fit is obtained using the ellipse equation, the general form of which is given by
\begin{ceqn}
\begin{equation}
\frac{(x-h)^2}{a^2} + \frac{(y-k)^2}{b^2} = 1
\end{equation}
\end{ceqn}
The following plot depicts the fit obtained from applying the ellipse equation to our dataset \cite{preincreasing}:
\begin{figure}[h!]
\centering
\includegraphics[width=0.6\textwidth]{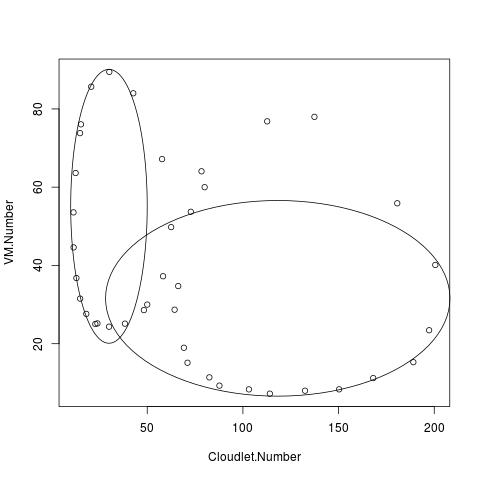}
\caption{In cloudsim jobs is cloudlets. Fitting VM and cloudlet dynamics data obtained from cloudsim simulation: We used least squares to predict the center and major/minor axes of the fitted ellipse. The fitted equation predicts the VM population against a given cloudlet population with reasonable degree of accuracy. This embellishes our model by providing predicting power.}
\label{Ellipse_fit}
\centering
\end{figure}
The following equations have been used to find the ellipses: 
\begin{ceqn}
\begin{equation}
\frac{(x-29.98445)^2}{20^2} + \frac{(y-55.116337)^2}{35^2} = 1
\end{equation}
\end{ceqn}
\begin{ceqn}
\begin{equation}
\frac{(x-118.21488)^2}{90^2} + \frac{(y-31.59861)^2}{25^2} = 1
\end{equation}
\end{ceqn}
\section*{Implementation in R}

We have used the ellipse package in R to generate the ellipse given the parameters $h, k, a, b$. The plotrix package is used to plot the ellipse and estimate the parameters with 95\% confidence. 
The function 'ellipse' (defines in the ellipse package) allows us to draw a two-dimensional ellipse that traces a bivariate normal density contour for a given mean vector, covariance matrix, and probabilistic proximity of the points lying on the ellipse trace. Using this function, we have computed the center of the ellipse with a reasonable degree of precision. We then computed the distance of the closest point and the furthest point from the dataset and set these values as $b$ and $a$ (major/minor axes of the ellipse, tweaking these values would generate a family of concentric ellipses which would help explain the elastic behavior of the model).\newline
where, $a$ = Length of longitudinal axis and 
$b$ = Length of transverse axis.\newline
Finally the ellipse that fits the curve is drawn using the 'plotrix' library. Let us consider the following example.\\
On giving an input $x$ (Cloudlet.Number)=38.34 to the first ellipse equation, we obtain $y$ = 23.317108996715966 where expected output is $y$ = 25.13. This gives us an accuracy of about $92.98$ \%.
\newline
Similarly on giving an input $x$ (Cloudlet.Number)=150.29 to the second ellipse equation, we get $y$ = 8.240179917046074 where expected output is $y=8.34$. This gives us an accuracy of about $98.8$ \%. These two samples, randomly selected from the pool of simulated data, are sufficient to testify the goodness of fit of the predictive model. We may forecast the required VM population using the fit when the cloudlet population is known.

\subsection{Interpretation of Resource-Demand elasticity:} 
We generated multiple traces of different ellipses centered around the same point (using the fitted equation and manipulating the parameters). We obtained different equations and subsequently different sets of $x$ (predator) and $y$ (prey) values -- this is the range of VM-jobs values that will help maintain the status quo i.e steady state (shape of the fitted curve is an ellipse) and provide a physical interpretation of elasticity without disturbing the equilibrium (since data was generated from the allocation algorithm specifically written to handle elasticity in service provisioning); also, we observe evidence of concavity from the scatter plot of the simulation, at least for a smaller range of jobs numbers. It might imply that when the jobs number varies between 50 and 100, the VM number rises significantly in the beginning and then falls sharply. Note that, for a set of jobs values, there is also possibility of attaining multiple elasticity that makes the choice less precise than one would expect under automatic allocation principle. Overall, except for a few outliers values of the elasticity thus arrived at, we do encounter concentric ellipses over a wider range of jobs
numbers. From a policy point of view, one could argue that the choice of jobs do offer a specific zone within which the VM numbers are expected to operate, making the problem of jobs choice less vexing. This is indeed a step forward in the analysis of this relationship, because previous attention to choice of jobs and the range of elasticity it varies within, have not been able to generate the elliptical path as we offer here. Therefore, if the allocation problem does not automatically guarantee presence of a steady state accompanied by small perturbations in the neighborhood, the choice of jobs numbers as part of intervention could ensure that under most circumstances.         
\par More importantly, our conjectures are not bounded by a specific structure (here, elliptical) generated from the fitted data. Let us assume that new experiments conducted by varying the parameters of the proposed ALVEC model generate new data points , which after fitting assume the shape of another conic section. Since our main purpose here is to provide a meaningful distribution, any other pattern, converging to or diverging from the steady state could be equally appealing to the decision problem. In fact, a conformal transformation converts a circle to an ellipse and inverse transform does the contrary. The transformation in shape, if at all happens, is a further evidence of elastic behavior in regions other than the equilibrium i.e steady state (ref Fig.4 where the equilibrium and non-equilibrium regions are clearly identified).
\par Furthermore, family of concentric ellipses satisfying the constrains of simulation parameters are generated. This pertains to the elasticity of the ellipses (fitted curve for jobs and resource relationship) and we found the following:
For ellipse 1: ${(x-29.98445)^2}/{20^2} + {(y-55.116337)^2}/{35^2} = 1$, the largest ellipse that can be formed has $a=29.98445$ and $b =55.116337$ and the smallest ellipse that can be formed has $a=0$ and $b =0$. \\
For ellipse 2: ${(x-118.21488)^2}/{90^2} + {(y-31.59861)^2}/{25^2} =1$, the largest ellipse that can be formed has $a=118.21488$ and $b =31.59861$ and the smallest ellipse that can be fitted are for $a=0$ and $b =0$.\\
\textit{Remark: The range of values for y are between $55.116337$ and $0$ (ellipse 1) and $31.59861$ and $0$ (ellipse 2); 
and for $x$, the range of values are between $29.98445$ and $0$ (ellipse 1), $118.21488$ and $0$ (ellipse 2). Ellipse 1 describes the scenario of prey increasing-predator decreasing and ellipse 2 describes the scenario of predator increasing-prey decreasing. It is well understood at this point that, predator signifies population of jobs (Cloudlets in CloudSim) and prey signifies population of resources (VM's in CloudSim). We plotted the original ellipse, accomplishing 95\% confidence in fitting the relationship between jobs and resources (ref to Fig. \ref{Ellipse_fit}). The family of concentric ellipses (ref to Fig. 21), generated by manipulating the major and minor axes of the original ellipse, provide us an extended range of $x, y $ values i.e. resource-job pairs that accommodate the elasticity restrictions. In summary, as long as the job population (demands) is within the range of $0-200.55$, we could predict the load/resource requirement with reasonable accuracy. The simulated data set may be found here \cite{preyincreasing} \cite{preydecreasing} \cite{SimData}.}
 \begin{figure*}[h!]
  \centering \includegraphics[width=3.0in]{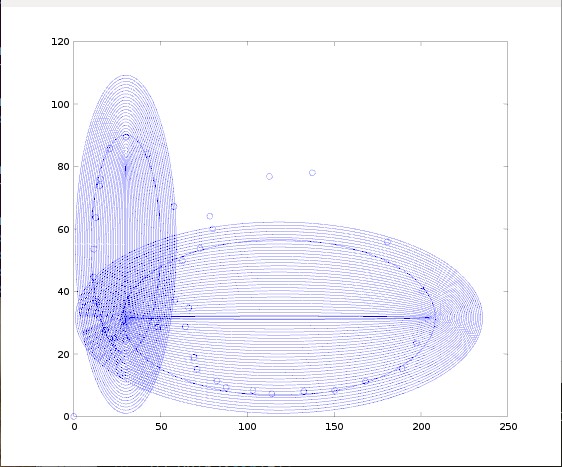}
  \caption{This is a family of concentric ellipses describing range of values of demand and resources. These ellipses are transformations in shape from the original Fig.\ref{Ellipse_fit}. The trace of $x$ and $y$ values indicate adaptive resource allocation against changing job demands. Our predictive algorithm adapts to elasticity in demand and allocation with good precision.}
\end{figure*}
 
As long as the fitted ellipse adheres to 95\% confidence restriction, we can be certain about predicting the resource matching jobs or demands within a range of values for number of jobs. In our case, the range of job demands is between $0-200.55$ (non integer values will be rounded off to the nearest integer) in order to maintain the prediction accuracy for resource requirements in a dynamic scenario. Clearly, in the absence of any demand ($x=0$), there is no resource requirement ($y=0$).
\par The last section shall summarize the wholesome contributions and lasting impact our work could bring about.  We note that, while many published manuscripts handled different performance issues of cloud, a comprehensive understanding, research and documentation on elasticity management is lacking or not convincing enough. Most of the allocation and management policies, although, novel are supervised and accrue overhead over time.

\section{Discussion and CONCLUSION}\label{CONCLUSION}
The Lotka-Volterra model provides us with a mathematical property known as limit cycles which is described in contour portraits or phase portraits of the system. Limit cycle describes a qualitative limit for the stability of a system. Parameters of a system are differed such that the system grows out of stability and difference acquired by the parameters is measured to state the domain of stability. This has direct application in understanding the stability of a web-server with incoming requests. Limit cycle of a system along with the rate of incoming requests can help us understand the bounds of the system. As the proposed model is based on the dynamic interaction between the VM and the cloudlet, our contribution addresses elasticity for highly volatile jobs or demands. ALVEC helps gauge the optimum level of elasticity. Time shared algorithm is another contribution. We implemented the experimental simulation in CloudSim. The proposed Lotka-Volterra model has the existing cloudlet time sharing algorithm on CloudSim and takes advantage of the predator-prey equation. Without externally defining any dynamic allocation scheduling algorithm, the improved time-shared algorithm decides the VM occupancy before submitting the incoming cloudlet request to VM. Other two contributions of this paper are Improvement on Quality of Services and minimization of SLA (Service Level Agreement) violation. The simulation in the CloudSim reveals that the number of future VM has not increased or decreased as per predefined static allocation rule. On the contrary, the model decides on the number of VM's "on the fly". The QoS metrics which include throughput, response time, etc show that the proposed model is more suitable to address each Quality metric. SLA defines the terms and conditions among two parties as a basis for measuring agreed quality of service standards and optimization between both Cloud provider and allocated job. Our contributed work quantifies the SLA violation parameter where non-fulfillment of services should be penalized. The proposed model shows significant reduction in SLA violation, i.e low penalty for not honoring SLA. This indirectly increases the profitability. This may have a disruptive impact by evolving a business model for the small and medium scale enterprises in Cloud business.  An analysis has been done from the economic point on how the entry-barrier challenge can be addressed for the new Cloud service providers.
It can be intuited from above that, a dynamic environment like Cloud follows lower dimensional chaos (Non linear dynamics). The motive was to bring about the cloud parameters under different situations and model them using non-linear dynamics. The parameters were calculated at the boundary conditions using a Java based simulation platform called CloudSim. The dynamic creation of cloudlets allowed us to vary the number of jobs and the resources of the Cloud to create a system which was able to showcase all possible scenarios. Lotka-Volterra model is suited as the best model to describe the Cloud parameters to reasonable accuracy. Phase portraits are used to determine over-utilization of resources leading the whole system into instability. A phase portrait is plotted repeatedly by using the Cloud data. When there is a sign of instability in the system, more resources can be added to bring the system into equilibrium state. 
We conclude by highlighting the strengths of our contribution.

\begin{itemize}
\item stability implies proportional change on VM based on changes of demand in jobs controlled by LV model. These differential equations are not affected by stochastic uncertainty. Therefore, the ballpark estimates of VM's based on jobs are achieved reliably and efficiently. The interpretation of stability between VM's and jobs is adroitly exploited in our work (Please refer to the subsection Prey-Predator stability).

\item The difference in modeling elasticity (our approach) from other approaches available in the literature needs to be highlighted. VM's are added in most approaches in supervised fashion whereas we allocate VM's governed by the underlying LV model without undermining the utilization threshold. This is an unsupervised approach. An unsupervised approach implies accomplishing parameter tuning by LV to control VM population (prey) corresponding to the cloudlet population (predator).
\item Our Cloudlet-timeshared algorithm performs better than existing timeshared algorithm implemented in Cloudsim.
 
\item We exploited the predictive analytics from the simulation of our LV model to estimate approximate VM population against the demand of jobs. Least Square fitting was applied to arrive at this estimate. As prediction accuracy is reasonable, there is lesser scope for overloading of systems.

\item Improvement in SLA violation was accomplished. Other QoS metrics are found to be better than  existing elastic work.

\item The technological innovation suggested by our model should encourage further competition in a market controlled by the big players.

\item VM allocation is accomplished by the proposed model, truly dynamic in fashion circumventing overheads or look up tables. Appropriate improvisations were effected in the basic cloudsim setting to accommodate dynamic behavior.

\item Adaptation and implementation of herding behavior is never implemented with strong intersection from population biology established for the first time. This is the cornerstone of our contribution.

\item We have shown improved utilization compared to other models. This, coupled with SLA violation minimization makes a strong case for resource optimality accomplished by ALVEC.
\par We implemented an optimization model for Cloud data centers so that no resource should be under-provisioned / over-provisioned. The greatest strength of the proposed model is its ability to handle jobs by dynamically deciding the scaling number and providing the input to the VM provisioning process. Our method doesn't pre-determine the number of VM's to be added. Every other resource allocation strategy uses a static allocation scheme. A pool of VM's, determined "on the fly" based on the population of jobs requested is assigned, instead of one VM being commissioned (which is usually the case). The pool of VM's may add to the cost since increased physical infrastructure would be needed. It would be interesting to investigate the equilibrium between incurred cost and improved SLA compliance due to our model. It is worth investigating energy awareness of our model in future. If resources have to be borrowed from another provider/pool, the model needs to be modified to accommodate such an event. However, the borrowing time shall add to the overhead. As the model decides the scaling number in a unsupervised fashion, it may suggest the number which is bigger than the static way of determining the scaling number. Therefore it may increase the cost of VM assignment. We haven't considered this possibility in our approach and we shall explore this as future work.
\par We note that, average RAM utilization is 44.17\% and average bandwidth utilization is 59.12\%. It is pertinent to observe two salient points:
\begin{itemize}
\item The methods compared with ALVEC did not report utilization of physical resources.
\item ALVEC did not encounter any resource failure (VMFailure in CloudSim) when simulations runs were performed. The proposed model did not have to deal with a scenario where demands could not have been met due to scarcity of resources, within the set bounds (resources are not unlimited, trivially). This has been validated across multiple simulation runs and stands out as a key performance  benchmark. The claim may be validated by running the code uploaded, \cite{LV}.
\end{itemize}


\end{itemize}

\section*{References}


\end{document}